\documentclass[aps,prd,onecolumn,eqsecnum,amsmath,nofootinbib,preprintnumbers]{revtex4}

\usepackage[dvipsnames]{xcolor}
\usepackage{color,graphicx,float,subfigure,xcolor}
\usepackage{amsfonts,amssymb,theorem,mathrsfs,times}
\usepackage{bm}
\usepackage{multirow}
\usepackage{mathtools}
\usepackage{amsfonts,amssymb,theorem,mathrsfs}
\usepackage{dsfont}
\usepackage{setspace}
\usepackage{amsmath} 
\usepackage{graphicx}
\usepackage[colorlinks,linkcolor=blue,anchorcolor=blue,citecolor=blue]{hyperref}
\usepackage{ulem}

\begin{document}
\title{Virtual absorption modes of Schwarzschild-de Sitter spacetimes in semi-open systems}
\preprint{\hfill {\small {ICTS-USTC/PCFT-26-22}}}
\date{\today}

\author{Liang-Bi Wu$^{a\, ,c}$\footnote{e-mail address: liangbi@mail.ustc.edu.cn}}

\author{Yu-Sen Zhou$^a$\footnote{e-mail address: zhou\_ys@mail.ustc.edu.cn}}

\author{Zhe Yu$^{d}$\footnote{e-mail address: yuzhe@nbu.edu.cn}} 

\author{Ming-Fei Ji$^{a}$\footnote{e-mail address: jimingfei@mail.ustc.edu.cn}}

\author{Li-Ming Cao$^{a\, ,b}$\footnote{e-mail address: caolm@ustc.edu.cn (corresponding author)}} 

\affiliation{${}^a$Interdisciplinary Center for Theoretical Study and Department of Modern Physics,\\
University of Science and Technology of China, Hefei, Anhui 230026, China}

\affiliation{${}^b$Peng Huanwu Center for Fundamental Theory, Hefei, Anhui 230026, China}

\affiliation{${}^c$School of Fundamental Physics and Mathematical Sciences, Hangzhou Institute for Advanced Study, UCAS, Hangzhou 310024, China}

\affiliation{${}^d$Institute of Fundamental Physics and Quantum Technology, Ningbo University, Ningbo 315211, China}

\begin{abstract}
We present a study of virtual absorption modes (VAMs) in Schwarzschild-de Sitter (SdS) spacetime under semi-open boundary conditions, where the VAMs correspond to total transmission modes (TTMs) with the reflection amplitude being vanishing. Our numerical analysis reveals that as the reflectivity $|\mathcal{K}|$ decreases, the imaginary parts of VAM spectra systematically decrease, with each overtone exhibiting a critical reflectivity at which $\text{Im}(\omega_{\text{VAM}})=0$. Using simulations based on spectral collocation methods, we demonstrate that excitation precisely at a VAM spectrum leads to a virtual absorption process. These results establish VAMs as the spectral signatures of virtual absorption processes for exotic compact objects (ECOs).
\end{abstract}

\maketitle

\section{Introduction}
In the characterization of the natural oscillations of black holes under perturbation, quasinormal modes (QNMs) emerge as a discrete spectrum of complex-valued frequencies. The real component denotes the oscillatory frequency, while the imaginary component signifies the damping rate. As distinctive spectral signatures of black holes, QNMs constitute a robust tool for testing the Kerr hypothesis and probing gravitational dynamics in the strong-field regime. This line of inquiry is known as black hole spectroscopy~\cite{Kokkotas:1999bd,Berti:2009kk,Konoplya:2011qq,Berti:2025hly,Cardoso:2025npr}.

Within the scattering theory, however, the QNMs represent just one prominent category of modes. Besides the QNMs, which correspond to the poles of the reflection and transmission amplitudes, there exists an additional family of modes referred to as the total transmission modes (TTMs). For a TTM spectrum, the reflection coefficient becomes zero, allowing an incident wave to pass through the effective potential barrier without any reflection~\cite{Berti:2004md,Andersson:1994tt}. These modes are characterized as complex-frequency solutions to the perturbation equation that exhibit the behavior of purely outgoing or purely ingoing waves in both boundaries. If it is $\text{TTM}_{\text{R}}$, both sides of the boundary are ingoing, and if it is $\text{TTM}_{\text{L}}$, both sides are outgoing. Here, we define an ingoing wave as a wave propagating toward decreasing radial coordinate, while an outgoing wave propagates toward increasing radial coordinate. From this definition, QNMs are defined by imposing ingoing at the event horizon and outgoing at the cosmological horizon or infinity. For $4$-dimensional Schwarzschild black holes, the algebraically special modes are effectively TTMs~\cite{Berti:2004md,Andersson:1994tt,MaassenvandenBrink:2000iwh}. When rotation is involved, the situation grows increasingly intricate. An asymptotic analysis of QNMs, TTMs, and total reflection modes (TRMs) was performed in~\cite{Keshet:2007be}. However, regarding the study of TTMs for rotating black holes, this will be credited to a series of works by Cook \textit{et al.}~\cite{Cook:2014cta,Cook:2016fge,Cook:2016ngj,Cook:2018ses,Cook:2022kbb}. TTMs of type-D black holes are studied in~\cite{Chen:2025sbz}. More recently, the notion of quasi-reflectionless scattering modes (RSMs) that are nothing but the TTMs are introduced in~\cite{Rosato:2025byu}. RSMs in asymmetric Damour-Solodukhin wormholes are investigated in~\cite{Qian:2025occ}. Just like using the pseudospectrum to study the (in)stability on QNMs (see e.g. \cite{Jaramillo:2020tuu} and references therein), the exploration of the pseudospectrum in TTMs is presented in~\cite{Zhou:2025xdo}. 

TTMs with positive imaginary parts warrant inclusion in the analysis, in contrast to the QNM framework where such a feature would imply an unstable system. Such a TTM can be selectively stimulated by an incoming waveform designed to grow exponentially in time, the rate of which is precisely set by the mode's imaginary part. In this scenario, the emergence of the reflected waveform is deferred until the exponential amplification is inevitably halted by either computational constraints or experimental boundaries. As a result, the whole black hole spacetime effectively becomes a perfect absorber during scattering~\cite{Tuncer:2025dnp}. Therefore, TTMs are also called the virtual absorption modes (VAMs).  

Given that realistic black holes deviate from the idealized concept of perfect darkness, they are more accurately described as exotic compact objects (ECOs). For these systems of ECOs, the boundary conditions for the waveforms around the event horizon will be modified into $\Psi\sim \mathrm{e}^{-\mathrm{i}\omega(x-x_{\text{w}})}+\mathcal{K}(\omega)\mathrm{e}^{\mathrm{i}\omega (x-x_{\text{w}})}$ in the frequency domain. The ringdown of exotic compact objects is distinguished by the phenomenon of gravitational wave (GW) echoes which are widely studied~\cite{Mark:2017dnq,Cardoso:2019apo,Bueno:2017hyj,Cardoso:2016rao,Cao:2025qxt,Cardoso:2016oxy,Zhang:2025ygb,Yang:2025hqk,Rosato:2025byu,Rosato:2025lxb}. For high-dimensional black holes, the double peak structure of the effective potential naturally produces GW echoes without the reflective boundary conditions~\cite{Cao:2024sot}. Several formulations of reflectivity have been rigorously derived, encompassing the wormhole, Boltzmann, area quantization~\cite{Oshita:2019sat,Wang:2019rcf,Deppe:2024fdo}, which are summarized in~\cite{Berti:2025hly,Rosato:2025byu}. There are several studies on how boundary conditions affect the stability of QNM spectra~\cite{Solidoro:2024yxi,Oshita:2025ibu,Wu:2025wbp,Destounis:2025dck}.

According to Green's function theory, the ringdown waveforms of gravitational waves generated by any source can, in principle, be calculated once two independent homogeneous solutions called ``in'' and ``up'' solutions are known. If these two solutions can be obtained analytically, the difficulty of constructing high-precision waveforms is greatly reduced. The perturbation equation of the Schwarzschild-de Sitter (SdS) black hole is one of the equations that can be analytically solved~\cite{Arnaudo:2025uos,Arnaudo:2025kit,Wu:2025wbp}. This is because the singular points of such equation are all regular, and there are only four of them. This equation can be transformed into the form of the standard Heun's equation. Recently, the importance of Heun's function has received increasing attention, and the literature on studying QNMs based on this useful function has flourished~\cite{Hatsuda:2020sbn,Noda:2022zgk,Chen:2025sbz,Chen:2024rov,Xia:2025hwt,Li:2026zsg,Mi:2025fbt,Li:2025lgn,Jiang:2025mcj}.

In this study, we focus on perturbations of the Schwarzschild-de Sitter (SdS) spacetime in semi-open systems to study the VAM spectra via the Heun's functions. To be more specific, we show that how the VAM spectra of the SdS spacetime are affected by reflectivity $\mathcal{K}(\omega)$ being near the event horizon. Motivated by the recent work~\cite{Tuncer:2025dnp}, and associated studies in other fields of physics~\cite{Baranov:2017,Maddi:2025,Trainiti:2019,Zhang:2026jsc} on virtual absorption and total absorption phenomena, we can precisely design the initial conditions so that the waveform is completely absorbed by the absorber composed of the reflective wall and the effective potential. The boundary conditions we use here are no longer Dirichlet boundary conditions which is used in~\cite{Tuncer:2025dnp}.

The remainder of this work is organized as follows. In Sec. \ref{sec: set ups}, we give a brief view on solving the perturbation equation of SdS black hole by the Heun's function~\cite{Wu:2025wbp}. In Sec. \ref{sec: VAMs}, we give the condition of the VAM spectra as the reflectivity exists, and the migration of the VAM spectra with respect to the reflectivity, in which the reflectivity is real. The eigenfunctions related to VAMs are also demonstrated. The numerical experiment of virtual absorption is presented in Sec. \ref{sec: simulation}. Sec. \ref{conclusions} is the conclusions and discussion. In Appendix \ref{C_functions}, the amplitude functions $C_{\text{in}}(\omega)$ and $C_{\text{out}}(\omega)$ of the ``down'' solution are work out. In Appendix \ref{numerical_method_TD}, we give the numerical method of solving the wave equation, which is based on the pseudospectral method. In Appendix \ref{Integral_method}, we give the numerical method of the integral on subinterval for the Chebyshev-Lobatto grids. Numerical resolution test is performed in Appendix \ref{Numerical_convergence_test}.

\section{Set ups}\label{sec: set ups}
The $4$-dimensional Schwarzschild-de Sitter (SdS) black hole is the solution of Einstein's equations. It describes a black hole in a de-Sitter background. The SdS black hole is completely specified by two parameters, namely $M$ and $\Lambda>0$. In a static coordinate, SdS black hole is described by the metric
\begin{eqnarray}\label{metric_and_function_f}
    \mathrm{d}s^2=-f(r)\mathrm{d}t^2+\frac{\mathrm{d}r^2}{f(r)}+r^2(\mathrm{d}\theta^2+\sin^2\theta\mathrm{d}\phi^2)\, ,\quad f(r)=1-\frac{2M}{r}-\frac{\Lambda r^2}{3}=-\Lambda\frac{(r-r_\text{c})(r-r_\text{e})(r-r_n)}{3r}\, ,
\end{eqnarray}
where the function $f(r)$ has been parameterized by the event horizon $r_\text{e}$ and the cosmological horizon $r_\text{c}$, and $r_n$ is the third real root of $f(r)$ with $r_n=-(r_\text{e}+r_\text{c})<0$. From Eq. (\ref{metric_and_function_f}), one can directly derive the relation between the cosmological constant $\Lambda>0$ and the black hole mass $M$ and horizons via~\cite{Konoplya:2022xid,Zhou:2025xta}
\begin{eqnarray}\label{mass_and_cosmological_constant}
    M=\frac{r_\text{c}r_\text{e}(r_\text{c}+r_\text{e})}{2(r_\text{c}^2+r_\text{c}r_\text{e}+r_\text{e}^2)}\, ,\quad \text{and}\quad
    \Lambda=\frac{3}{r_\text{e}^2+r_\text{e}r_\text{c}+r_\text{c}^2}\, .
\end{eqnarray}
The master perturbation equation in the time domain is
\begin{eqnarray}\label{master_equation_time_domain}
    \Big[\frac{\partial^2}{\partial t^2}-\frac{\partial^2}{\partial x^2}+V_s(x)\Big]\Psi(t,x)=0\, ,\quad \mathrm{d}x=\frac{\mathrm{d}r}{f(r)}\, ,
\end{eqnarray}
with the tortoise coordinate $x$ being 
\begin{eqnarray}\label{tortoise_coordinates_integration}
    x=\frac{3r_\text{e}}{\Lambda(r_\text{c}-r_\text{e})(r_\text{e}-r_n)}\ln\Big|1-\frac{r_\text{e}}{r}\Big|+\frac{3r_\text{c}}{\Lambda(r_\text{e}-r_\text{c})(r_\text{c}-r_n)}\ln\Big|1-\frac{r_\text{c}}{r}\Big|+\frac{3r_n}{\Lambda(r_\text{e}-r_n)(r_n-r_\text{c})}\ln\Big|1-\frac{r_n}{r}\Big|\, .
\end{eqnarray}
Here, we consider three kinds of potentials whose expressions are~\cite{Zhidenko:2003wq,Crispino:2013pya,Arnaudo:2025kit}
\begin{eqnarray}\label{potentials}
    V_{s}(r)=f(r)\Big[\frac{\ell(\ell+1)}{r^2}+(1-s)\frac{f^{\prime}(r)}{r}+\delta_{s,2}f^{\prime\prime}(r)+\frac{2\Lambda}{3}\delta_{s,0}\Big]\, ,\quad \text{for} \quad s=0\, ,1\, ,2\, ,
\end{eqnarray}
where $s=0,1,2$ indicate scalar, electromagnetic and gravitational axial perturbations, respectively, and $\delta_{s,2}$, $\delta_{s,0}$ stand for the Kronecker delta. It should be emphasized that the parameter $s$ used in Eq. (\ref{potentials}) serves as a label in the unified potential representation. It should not be interpreted as the spin weight appearing in the Newman-Penrose or Teukolsky formalism. Regarding the scalar perturbation $s=0$, it should be noted that the case discussed here is the conformally coupled massless scalar perturbation. In the frequency domain, with the time part $\mathrm{e}^{-\mathrm{i}\omega t}$, the perturbation equation (\ref{master_equation_time_domain}) can be reduced to the following Schr\"{o}dinger wave-like equation
\begin{eqnarray}\label{master_equation_frequency_domain}
    \Big[\frac{\mathrm{d}^2}{\mathrm{d}x^2}+\omega^2-V_s(x
    )\Big]\Psi(x)=0\, .
\end{eqnarray}
For this homogeneous equation, there are some solutions that satisfy specific boundary conditions. The standard ``in'' and ``up'' solutions of Eq. (\ref{master_equation_frequency_domain}) are defined as
\begin{eqnarray}\label{in_solution}
    \Psi_{\text{in}}(x)=
    \left\{
    \begin{array}{l}
    \mathrm{e}^{-\mathrm{i}\omega x}\, ,\quad x\to-\infty\\
    A_{\text{in}}(\omega)\mathrm{e}^{-\mathrm{i}\omega x}+A_{\text{out}}(\omega)\mathrm{e}^{\mathrm{i}\omega x}\, ,\quad x\to+\infty
    \end{array}\right.\, ,
\end{eqnarray}
and 
\begin{eqnarray}\label{up_solution}
    \Psi_{\text{up}}(x)=
    \left\{
    \begin{array}{l}
    B_{\text{in}}(\omega)\mathrm{e}^{-\mathrm{i}\omega x}+B_{\text{out}}(\omega)e^{\mathrm{i}\omega x}\, ,\quad x\to-\infty\\
    \mathrm{e}^{\mathrm{i}\omega x}\, ,\quad x\to+\infty
    \end{array}\right.\, .
\end{eqnarray}
From the above two equations, we can see that the ``in'' solution is ingoing at event horizon, and the ``up'' solutions is outgoing at cosmological horizon. Here, another useful solution called ``down'' solution of Eq. (\ref{master_equation_frequency_domain}) is defined as
\begin{eqnarray}\label{down_solution}
    \Psi_{\text{down}}(x)=
    \left\{
    \begin{array}{l}
    C_{\text{in}}(\omega)\mathrm{e}^{-\mathrm{i}\omega x}+C_{\text{out}}(\omega)e^{\mathrm{i}\omega x}\, ,\quad x\to-\infty\\
    \mathrm{e}^{-\mathrm{i}\omega x}\, ,\quad x\to+\infty
    \end{array}\right.\, .
\end{eqnarray}
To ensure the uniqueness of the coefficients $A_{\text{in}}(\omega)$, $A_{\text{out}}(\omega)$, $B_{\text{in}}(\omega)$, $B_{\text{out}}(\omega)$, $C_{\text{in}}(\omega)$ and $C_{\text{out}}(\omega)$, it is emphasized that the equal signs in equations (\ref{in_solution}), (\ref{up_solution}) and (\ref{down_solution}) are true equal signs, not proportional notations. After a M\"{o}bius transformation and a scale transformation, Eq. (\ref{master_equation_frequency_domain}) becomes the standard Heun equation~\cite{Arnaudo:2025uos,Arnaudo:2025kit,Wu:2025wbp}. Subsequently, the above amplitude functions can be solved. The detailed expressions of $A_{\text{in}}(\omega)$, $A_{\text{out}}(\omega)$, $B_{\text{in}}(\omega)$ and $B_{\text{out}}(\omega)$ can be found in~\cite{Wu:2025wbp}. In addition, functions $C_{\text{in}}(\omega)$ and $C_{\text{out}}(\omega)$ can be obtained from $A_{\text{in}}(\omega)$, $A_{\text{out}}(\omega)$, $B_{\text{in}}(\omega)$ and $B_{\text{out}}(\omega)$ (see Appendix \ref{C_functions}).

Consider the semi-open problem in which a wall of reflectivity $\mathcal{K}(\omega)$ is located at $x_\text{w}\ll0$ such that $V_s(x_\text{w})$ is approximated as zero, and the solution $\Psi_s(x)$ at $x=x_\text{w}$ is a linear combination of ingoing and outgoing waves $\mathrm{e}^{\pm\mathrm{i}\omega x}$. For such a situation, the solution denoted by $\Psi_s(x)$ of Eq. (\ref{master_equation_frequency_domain}) obey the following boundary condition
\begin{eqnarray}\label{semi_open_conditions}
    \Psi_s(x)= \left\{
    \begin{array}{l}
    \mathrm{e}^{-\mathrm{i}\omega x}+\mathcal{K}(\omega)\mathrm{e}^{-2\mathrm{i}\omega x_\text{w}} \mathrm{e}^{\mathrm{i}\omega x}\, ,\quad x\to x_\text{w}\\
    S_{\text{in}}(\omega)\mathrm{e}^{-\mathrm{i}\omega x}+S_{\text{out}}(\omega)\mathrm{e}^{\mathrm{i}\omega x}\, ,\quad x\to+\infty
    \end{array}\right.\, .
\end{eqnarray}
The zeros of $S_{\text{in}}(\omega)$ are nothing but the QNM spectra for the semi-open problem.

% For the semi-open problem associated with the solution $\Psi_s(x)$, its scattering matrix $\mathbf{S}(\omega)$ is expressed as
% \begin{eqnarray}
%     \begin{bmatrix}
%         1\\
%         S_{\text{out}}(\omega)
%     \end{bmatrix}
%     =\mathbf{S}(\omega)\cdot
%     \begin{bmatrix}
%         \mathcal{K}(\omega)\mathrm{e}^{-2\mathrm{i}\omega x_\text{w}}\\
%         S_{\text{in}}(\omega)
%     \end{bmatrix}\, ,\quad 
%     \mathbf{S}(\omega)=
%     \begin{bmatrix}
%         \mathbf{S}_{11}(\omega) & \mathbf{S}_{12}(\omega)\\
%         \mathbf{S}_{21}(\omega) & \mathbf{S}_{22}(\omega)
%     \end{bmatrix}\, ,
% \end{eqnarray}
% where matrix elements of $\mathbf{S}(\omega)$ read
% \begin{eqnarray}
%     \mathbf{S}_{11}(\omega)=\frac{B_{\text{in}(\omega)}}{B_{\text{out}}(\omega)}\, ,\quad\mathbf{S}_{11}(\omega)=\frac{1}{B_{\text{out}}(\omega)}\, ,\quad \mathbf{S}_{21}(\omega)=\frac{1}{A_{\text{in}}(\omega)}\, ,\quad\mathbf{S}_{22}(\omega)=\frac{A_{\text{out}(\omega)}}{A_{\text{in}}(\omega)}\, .
% \end{eqnarray}

\section{Virtual absorption modes}\label{sec: VAMs}
The virtual absorption modes, which are also called the total transmission modes (TTMs), are defined by their asymptotic at the boundaries. For the open system, like the QNMs, the right TTM ($\text{TTM}_{\text{R}}$) and left TTM ($\text{TTM}_{\text{L}}$) are defined by choosing the different signs of the following
\begin{eqnarray}
    \Psi(x)\sim\left\{
    \begin{array}{l}
    \mathrm{e}^{\pm\mathrm{i}\omega x}\, ,\quad x\to -\infty\\
    \mathrm{e}^{\pm\mathrm{i}\omega x}\, ,\quad x\to+\infty
    \end{array}\right.\, ,
\end{eqnarray}
where the corresponding effective potential should be zero at the boundaries (event horizon and cosmological horizon or infinity). Specifically, QNMs are defined by the asymptotic behaviors $\mathrm{e}^{-\mathrm{i}\omega x}$ as $x \rightarrow -\infty$ and $\mathrm{e}^{+\mathrm{i}\omega x}$ as $x \rightarrow +\infty$. In contrast, right TTMs are characterized by the $\mathrm{e}^{-\mathrm{i}\omega x}$ branch at both boundaries, while left TTMs take the $\mathrm{e}^{+\mathrm{i}\omega x}$ branch at both. As for the semi-open system currently under consideration, it is naturally to consider the solution of Eq. (\ref{master_equation_frequency_domain}) satisfies 
\begin{eqnarray}\label{TTM_R_boundary_condition}
    \Psi_s(x)\sim\mathrm{e}^{-\mathrm{i}\omega x}\, ,\quad x\to+\infty\, .
\end{eqnarray}
From such boundary condition, it can be inferred that solution $\Psi_s(x)$ is ingoing at the cosmological horizon. Combining Eq. (\ref{semi_open_conditions}) and Eq. (\ref{TTM_R_boundary_condition}), for the semi-open problem, TTM spectra are derived by the condition $S_{\text{out}}(\omega)=0$, where the expression of $S_{\text{out}}(\omega)$ reads
\begin{eqnarray}
    S_{\text{out}}(\omega)=A_{\text{out}}(\omega)+\frac{\mathcal{K}(\omega)\mathrm{e}^{-2\mathrm{i}\omega x_\text{w}}}{A_{\text{in}}(\omega)}\Big[1-A_{\text{out}}(\omega)B_{\text{in}}(\omega)\Big]\, ,\quad \omega\in\mathbb{C}\, .
\end{eqnarray}
Note that the TTM spectra are also defined as the zeros of the reflection amplitude, $\mathcal{R}(\omega)\equiv S_{\text{out}}(\omega)/S_{\text{in}}(\omega)$, while QNM spectra are the poles of the reflection amplitude. At this point, we refer to TTMs $\omega$ as the virtual absorption modes (VAMs) which will be denoted by $\omega_{\text{VAM}}$ (The reason why it is called this name is mostly due to time-domain reasons.). Here, we always focus on the axial gravitational perturbation corresponding to the $s=2$ case of the unified potential (\ref{potentials}). The angular momentum parameter $\ell$ is set to be $2$ with the event horizon being fixed $r_{\text{e}}=1$. In addition, for simplicity, the reflectivity is assumed to be a frequency-independent constant, i.e.,  $\mathcal{K}(\omega)=\mathcal{K}$. As for the method to solve for the zeros of $S_{\text{out}}(\omega)$, we adopt an approach similar to that in~\cite{Wu:2025sbq,Wu:2025wbp}. First, densely sample points in the complex $\omega$ plane, and then get $\ln|S_{\text{out}}(\omega)|$. In this way, we can obtain approximate values of zeros which are the minimum points of $\ln|S_{\text{out}}(\omega)|$. For $\mathcal{K}=1$ or $\mathcal{K}=-1$, using these approximate values as initial guesses for \textit{FindRoot} in \textit{Mathematica}, accurate values of zeros can be obtained. The zeros for other $\mathcal{K}$ can be obtained by slowly varying $\mathcal{K}$ from $\mathcal{K}=1$ or $\mathcal{K}=-1$.

\begin{figure}[htbp]
    \centering
   \subfigure[]{\includegraphics[width=0.45\linewidth]{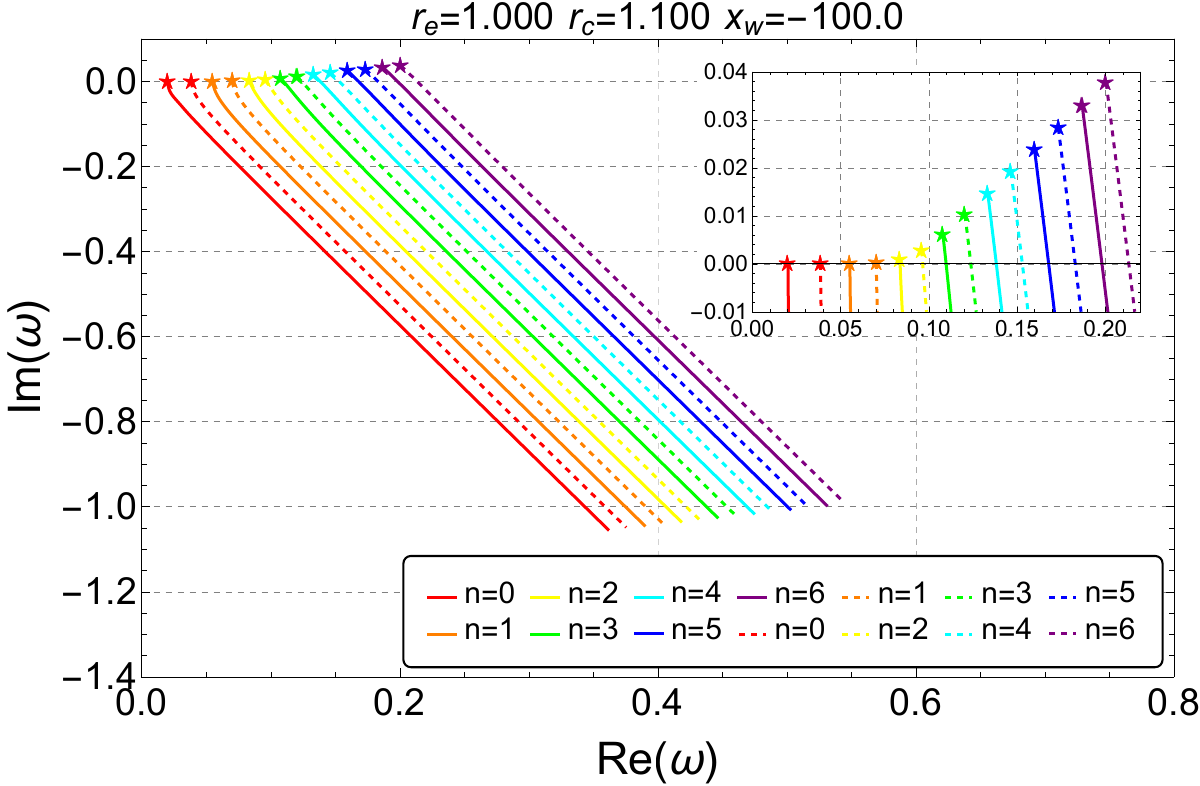}\label{TTM_Spectra_rc_1p1_a}}\hfill
   \subfigure[]{\includegraphics[width=0.45\linewidth]{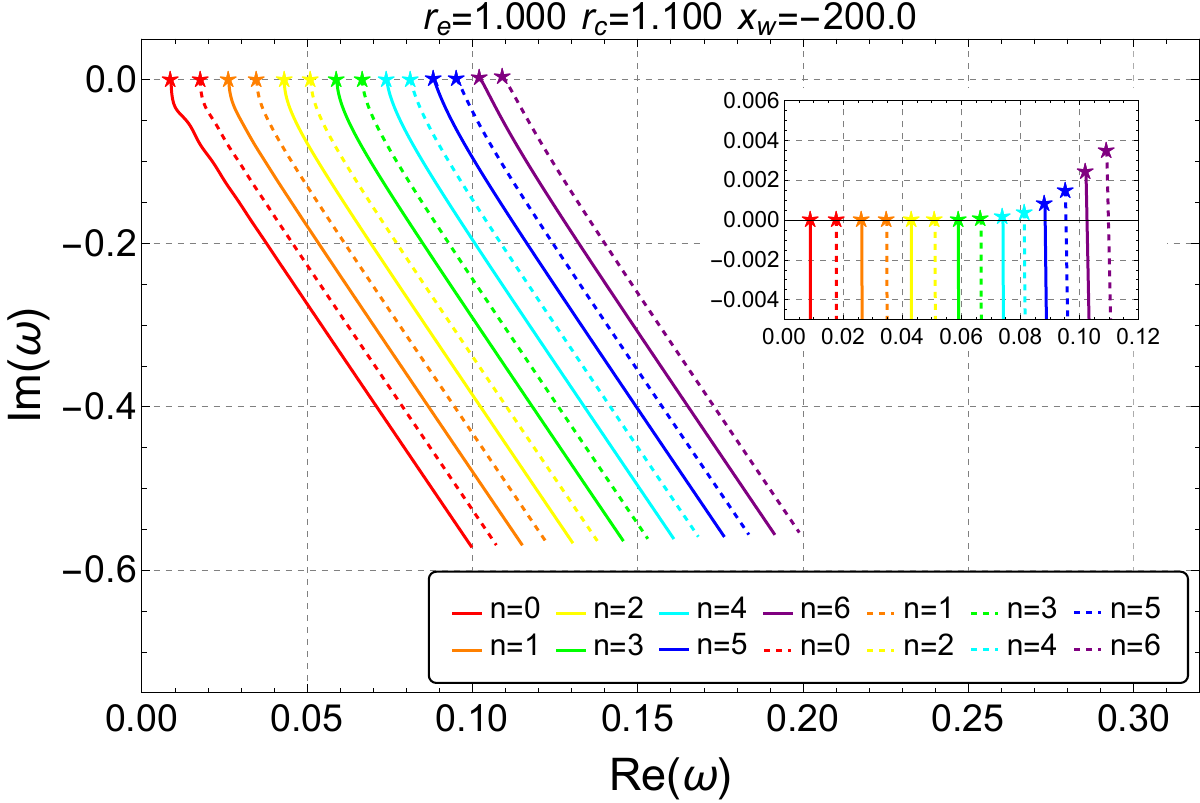}\label{TTM_Spectra_rc_1p1_b}}
   \subfigure[]{\includegraphics[width=0.45\linewidth]{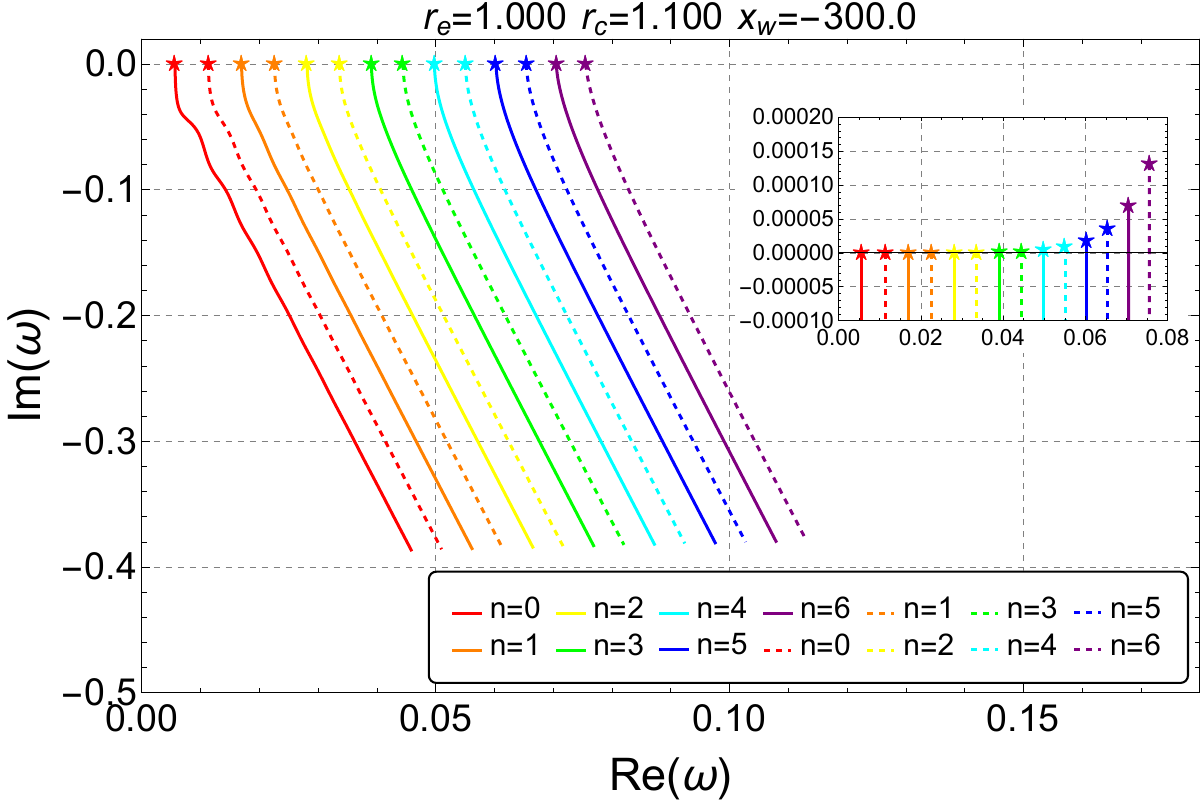}\label{TTM_Spectra_rc_1p1_c}}\hfill
   \subfigure[]{\includegraphics[width=0.45\linewidth]{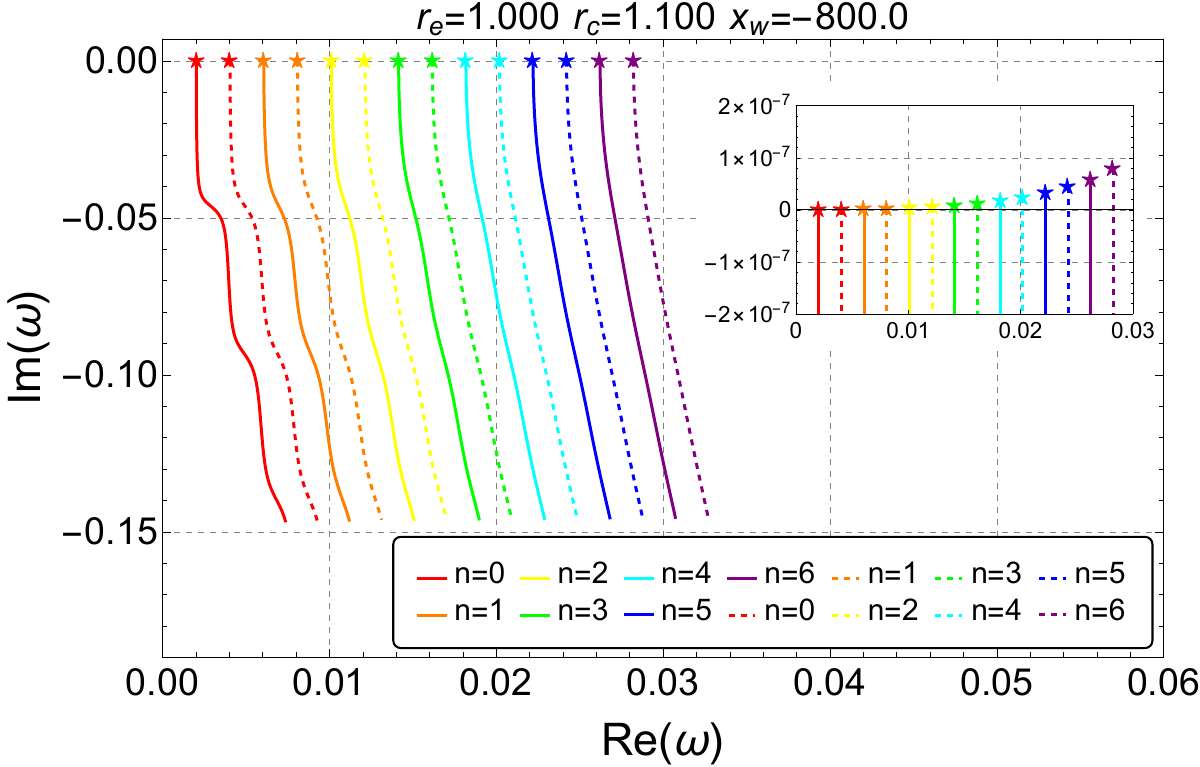}\label{TTM_Spectra_rc_1p1_d}}
    \caption{The migrations of the VAM spectra with $\mathcal{K}$ varying with $r_\text{c}=1.1$. For four panels, the positions of the reflective walls are $x_{\text{w}}=-100$, $x_{\text{w}}=-200$, $x_{\text{w}}=-300$, and $x_{\text{w}}=-800$, respectively. Different color lines represents different modes. Solid lines are for $\mathcal{K}>0$ while dashed lines are for $\mathcal{K}<0$.}
    \label{TTM_Spectra_rc_1p1}
\end{figure}

\begin{figure}[htbp]
    \centering
    \subfigure[]{\includegraphics[width=0.45\linewidth]{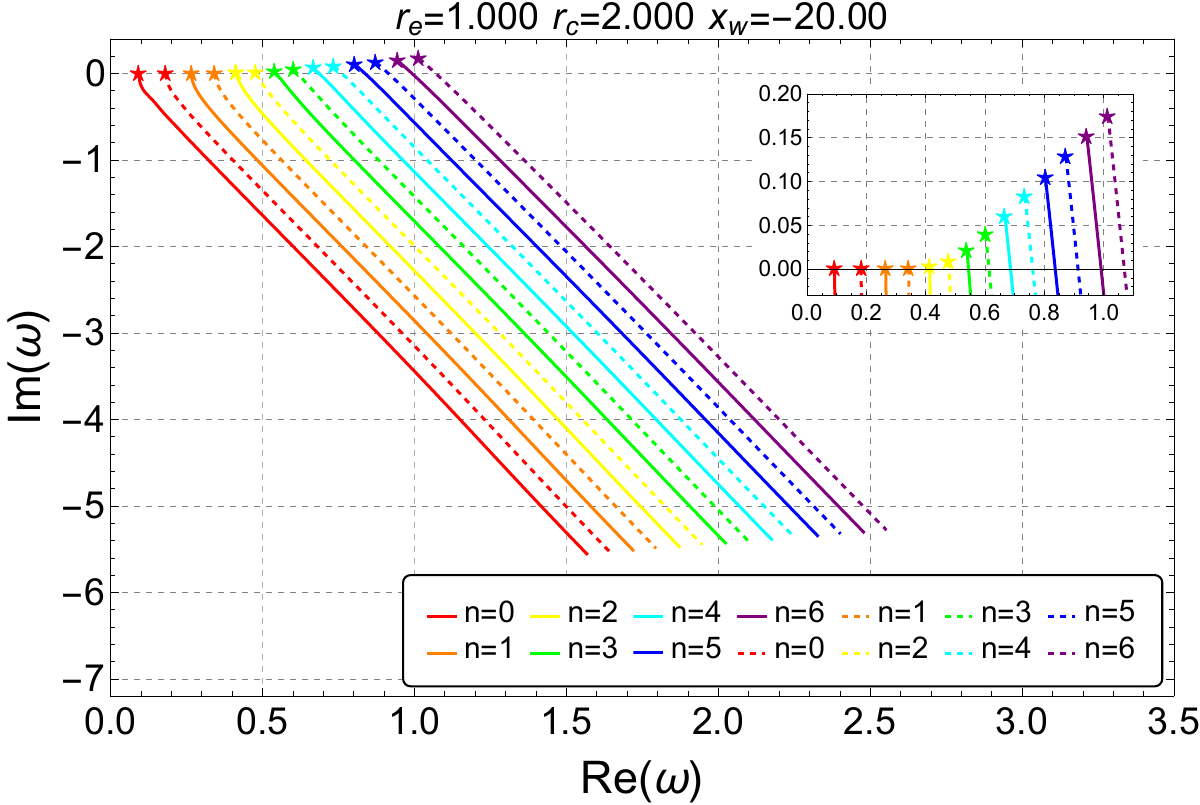}\label{TTM_Spectra_rc_2_a}}\hfill
    \subfigure[]{\includegraphics[width=0.45\linewidth]{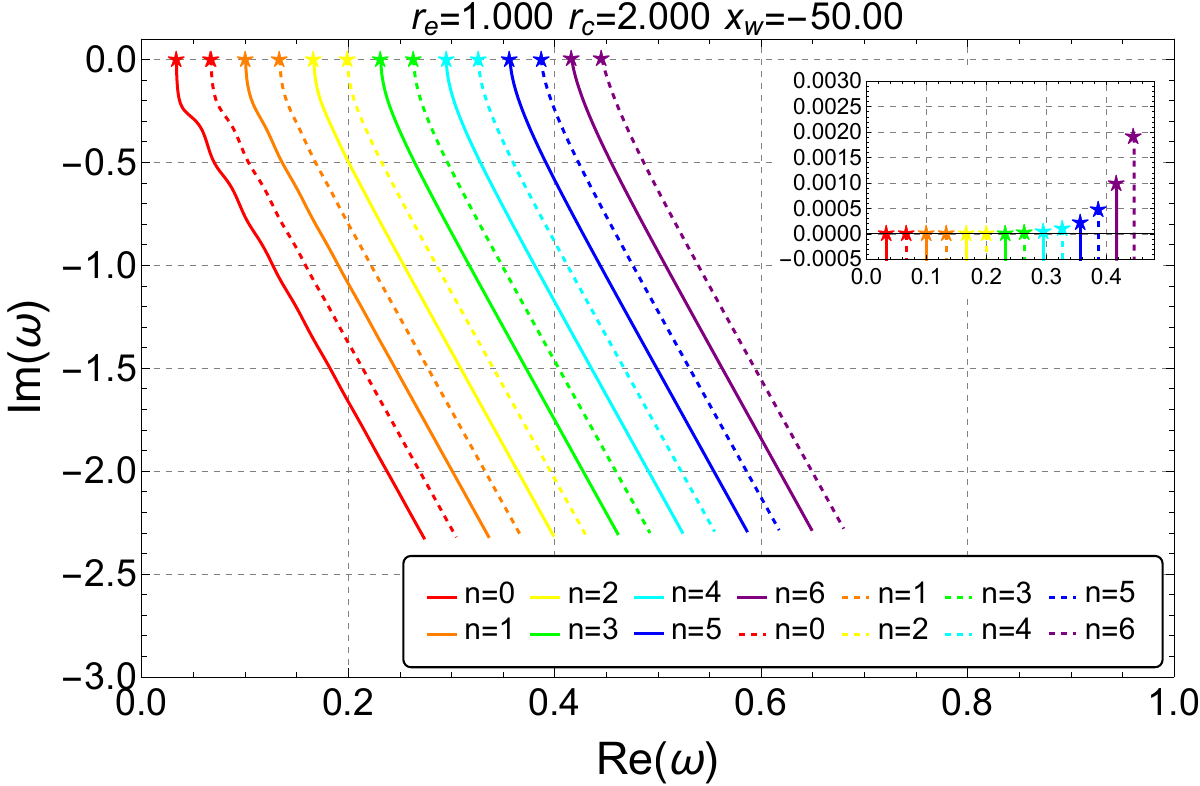}}
   \subfigure[]{\includegraphics[width=0.45\linewidth]{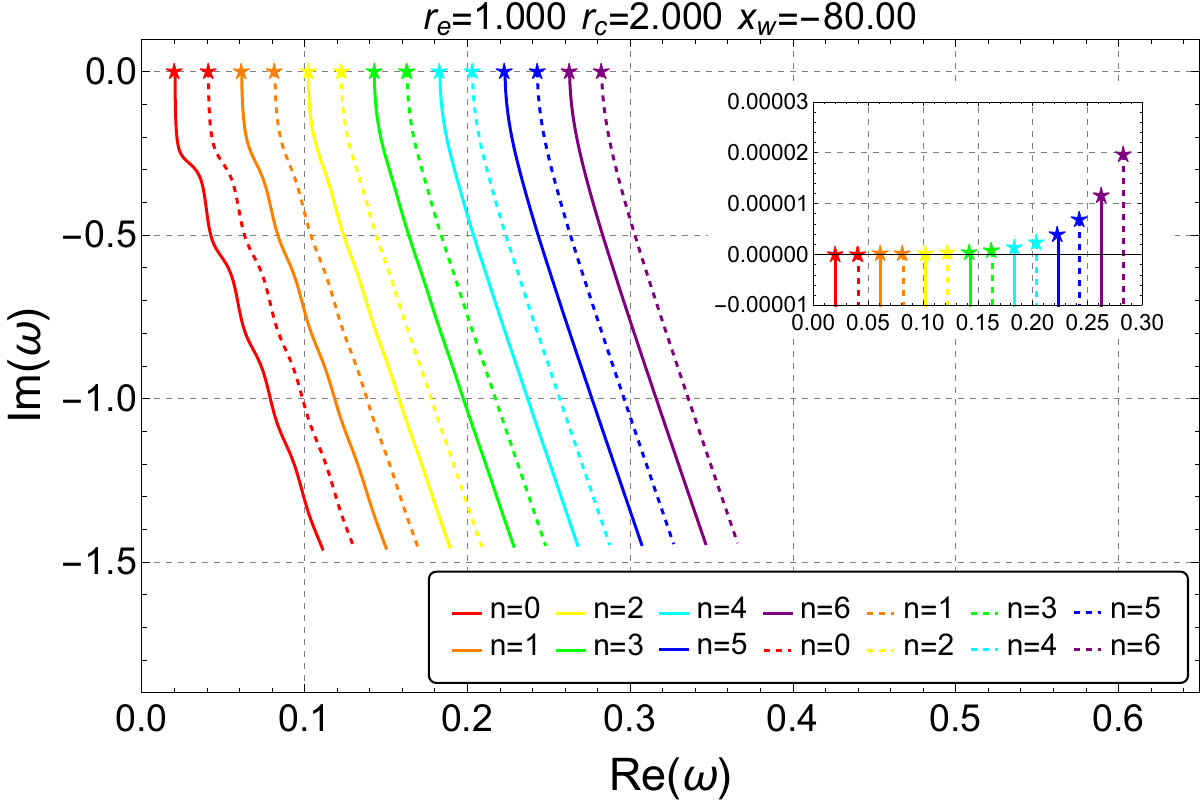}}\hfill
   \subfigure[]{\includegraphics[width=0.45\linewidth]{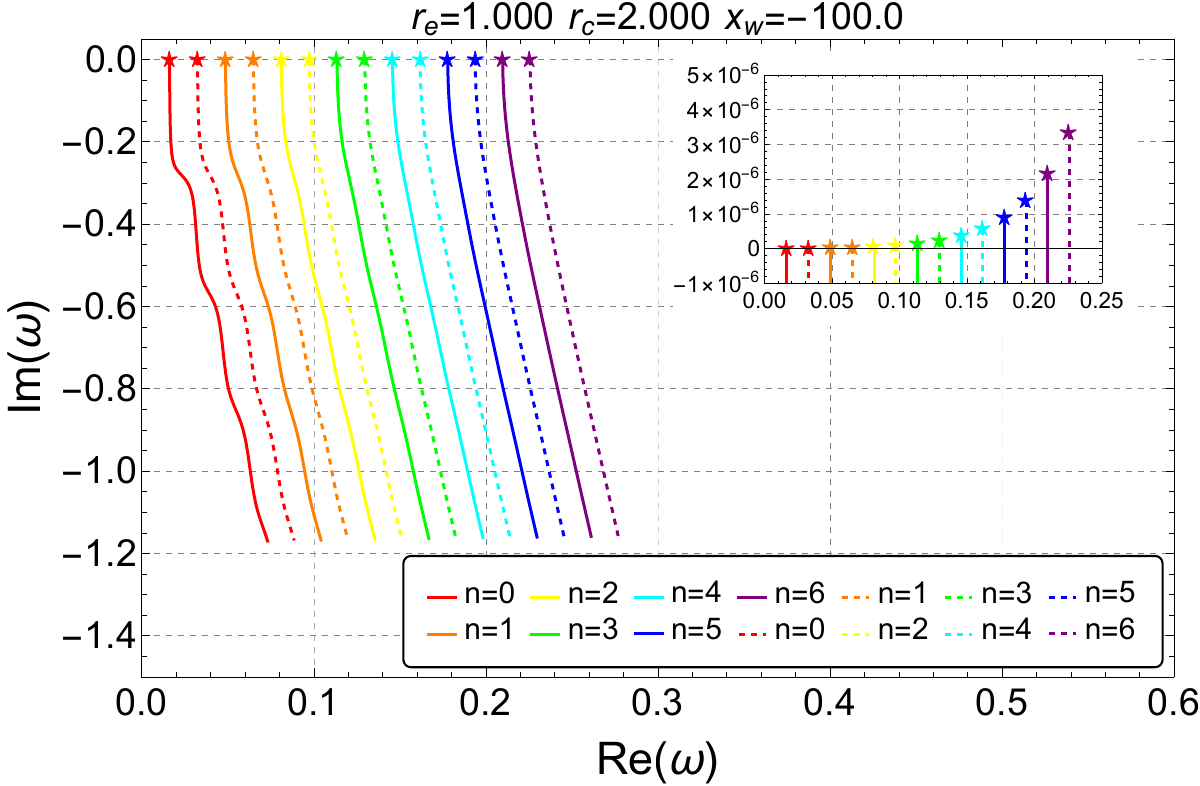}}
    \caption{The migrations of the VAM spectra with $\mathcal{K}$ varying with $r_\text{c}=2$. For four panels, the positions of the reflective walls are $x_{\text{w}}=-20$, $x_{\text{w}}=-50$, $x_{\text{w}}=-80$, and $x_{\text{w}}=-100$, respectively. Different color lines represents different modes. Solid lines are for $\mathcal{K}>0$ while dashed lines are for $\mathcal{K}<0$.}
    \label{TTM_Spectra_rc_2}
\end{figure}

In Fig. \ref{TTM_Spectra_rc_1p1} and Fig. \ref{TTM_Spectra_rc_2}, we show the virtual absorption modes with some typical parameters, where $r_{\text{c}}=1.1$ is chosen in Fig. \ref{TTM_Spectra_rc_1p1} and $r_{\text{c}}=2$ is chosen in Fig. \ref{TTM_Spectra_rc_2}. The solid lines in each panel represent the variations of the VAMs as $\mathcal{K}$ ranges over $[10^{-100}, 1]$. The dashed lines in each panel show the variations of the VAMs for $\mathcal{K}$ in the interval $[-1, -10^{-100}]$. Note that $\mathcal{K}=1$ yields the Neumann boundary condition, while $\mathcal{K}=-1$ yields the Dirichlet boundary condition. The stars in both figures correspond to the VAMs at $\mathcal{K}=1$ or $\mathcal{K}=-1$. The modes are labeled by the overtone number $n$, which is determined from the ordering of their imaginary parts at $\mathcal{K}=\pm1$. The same label is then assigned to the continuously connected spectral branch as $\mathcal{K}$ varies. In Fig. \ref{TTM_Spectra_rc_1p1} and Fig. \ref{TTM_Spectra_rc_2}, solid lines are for $\mathcal{K}>0$ while dashed lines are for $\mathcal{K}<0$.


This difference in $r_\text{c}$ alters the effective potential, thereby influencing the quantitative details of the VAM spectra. However, some qualitative characteristics do not change with the variation of $r_\text{c}$. From Fig. \ref{TTM_Spectra_rc_1p1} and Fig. \ref{TTM_Spectra_rc_2}, we can observe several characteristics of VAM spectra migration due to changes in $\mathcal{K}$. First, we can see that as $|\mathcal{K}|$ decreases, the modes consistently migrate towards regions of the complex plane characterized by more negative imaginary parts, namely, $-\text{Im}(\omega)$ becomes larger. Their trajectories will cross the real axis. For different modes, there always exists a critical $\mathcal{K}$ that causes the imaginary part of the VAM spectrum to vanish, namely $\text{Im}(\omega_{\text{VAM}})=0$, where these modes are called total absorption modes (TAMs)~\cite{Tuncer:2025dnp}. The real part also shifts, but the magnitude of this shift is generally smaller than the change in the imaginary part. Second, the lines of different colors are roughly arranged in parallel, and the spacing between them is also roughly the same. Such spacing will decrease as the reflective wall gets closer to the event horizon.

In Ref. \cite{Tuncer:2025dnp}, VAMs were found to coincide with the complex conjugates of QNMs in the case of Dirichlet boundary conditions $(\mathcal{K}=-1)$. In our present work, we focus on the virtual absorption modes of SdS spacetime in semi-open systems. For $\mathcal{K}=1$ which actually is the Neumann boundary condition, we check the results of TTMs and QNMs with $r_{\text{c}}=1.1$ and $x_{\text{w}}=-100$. We find that these VAMs are the complex conjugates of QNMs, as can be seen by comparing the QNM spectra in the left panel of Fig. 1 of Ref. \cite{Wu:2025wbp} (see asterisks therein) with the TTM spectra in panel (a) of Fig. 1 of the present manuscript. Attention should be paid to the difference in the convention of overtones between the two studies. The reason can be explained as follows. Since the frequency-domain radial equation has real coefficients, the complex conjugate of any radial solution is also a solution, with $\omega$ replaced by $\omega^{\star}$. In particular, complex conjugation maps the outgoing QNM behavior $\mathrm{e}^{+\mathrm{i}\omega_{\text{QNM}}x}$ at the cosmological horizon to the ingoing TTM behavior $\mathrm{e}^{-\mathrm{i}\omega_{\text{QNM}}^{\star}x}$. Moreover, the Dirichlet and Neumann conditions at the wall are frequency independent and invariant under complex conjugation. Therefore, the complex conjugate of a radial solution with frequency $\omega_{\text{QNM}}$ is a solution with frequency $\omega_{\text{QNM}}^{\star}$ and satisfies both the radial equation and the boundary conditions of TTMs. Alternatively, the same relation can be understood from the spherical symmetry. Since the radial equation depends on $\omega$ only through $\omega^{2}$, replacing $\omega$ by $-\omega$ leaves the radial equation unchanged. Under this replacement, the outgoing QNM condition at the cosmological horizon becomes the ingoing TTM condition. Thus, a QNM with frequency $\omega_{\text{QNM}}$ corresponds to a TTM with frequency $-\omega_{\text{QNM}}$. Moreover, in spherical symmetry case, the TTM spectra are symmetric under reflection about the imaginary axis. Then it also follows that $\omega_{\text{QNM}}^{\star}$ is a TTM spectrum. For other $\mathcal{K}$ with $|\mathcal{K}|<1$, the conclusion that TTMs and QNMs are complex conjugates of each other is not valid. This is because the boundary conditions at $x_{\text{w}}$ now depends on $\omega$.

\begin{figure}[htbp]
    \centering
    \subfigure[]{\includegraphics[width=0.31\linewidth]{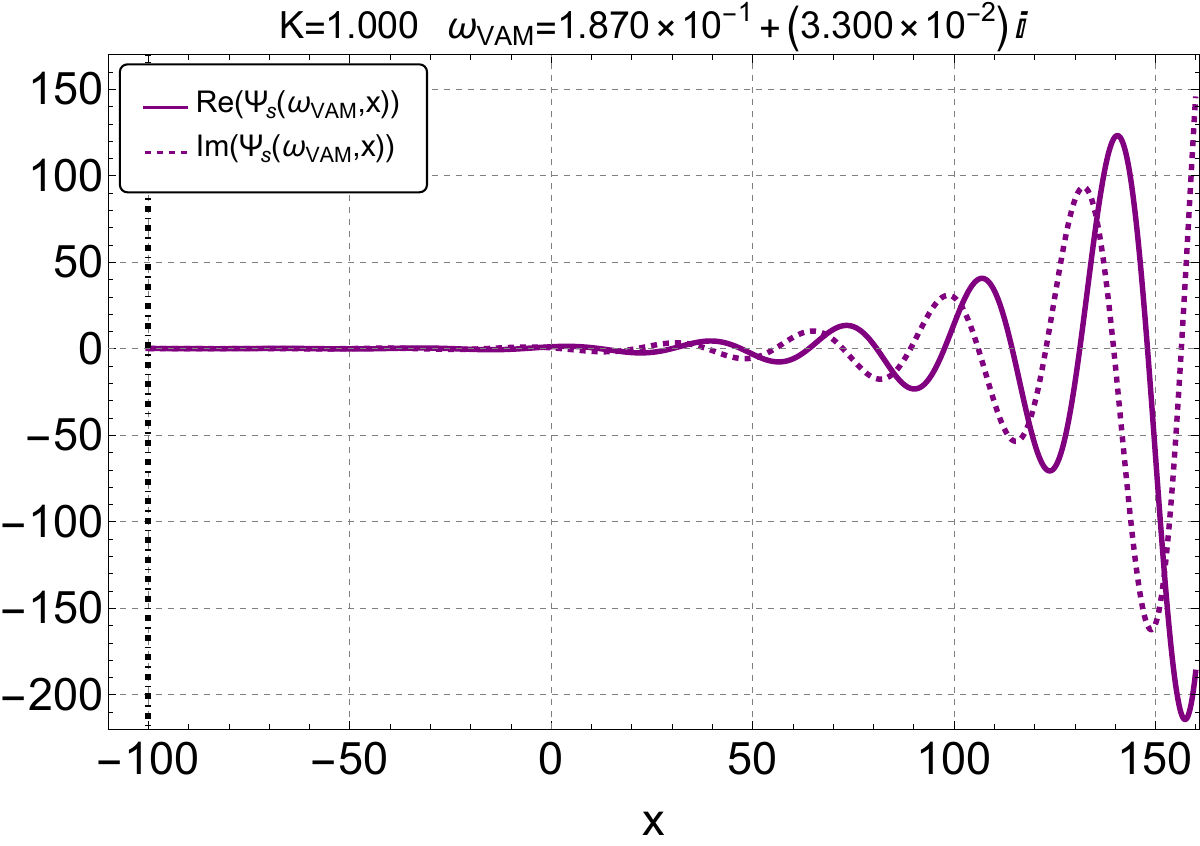}}\hfill
    \subfigure[]{\includegraphics[width=0.31\linewidth]{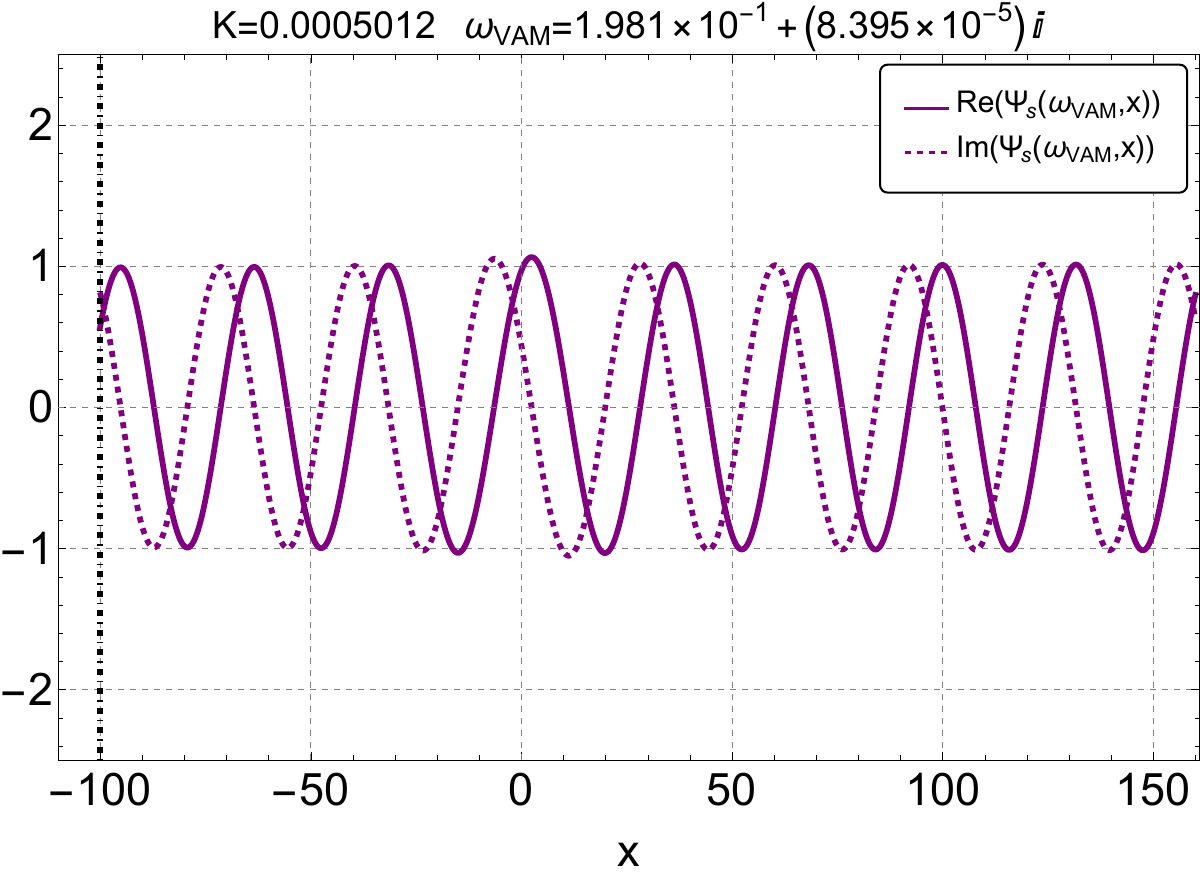}}\hfill
    \subfigure[]{\includegraphics[width=0.31\linewidth]{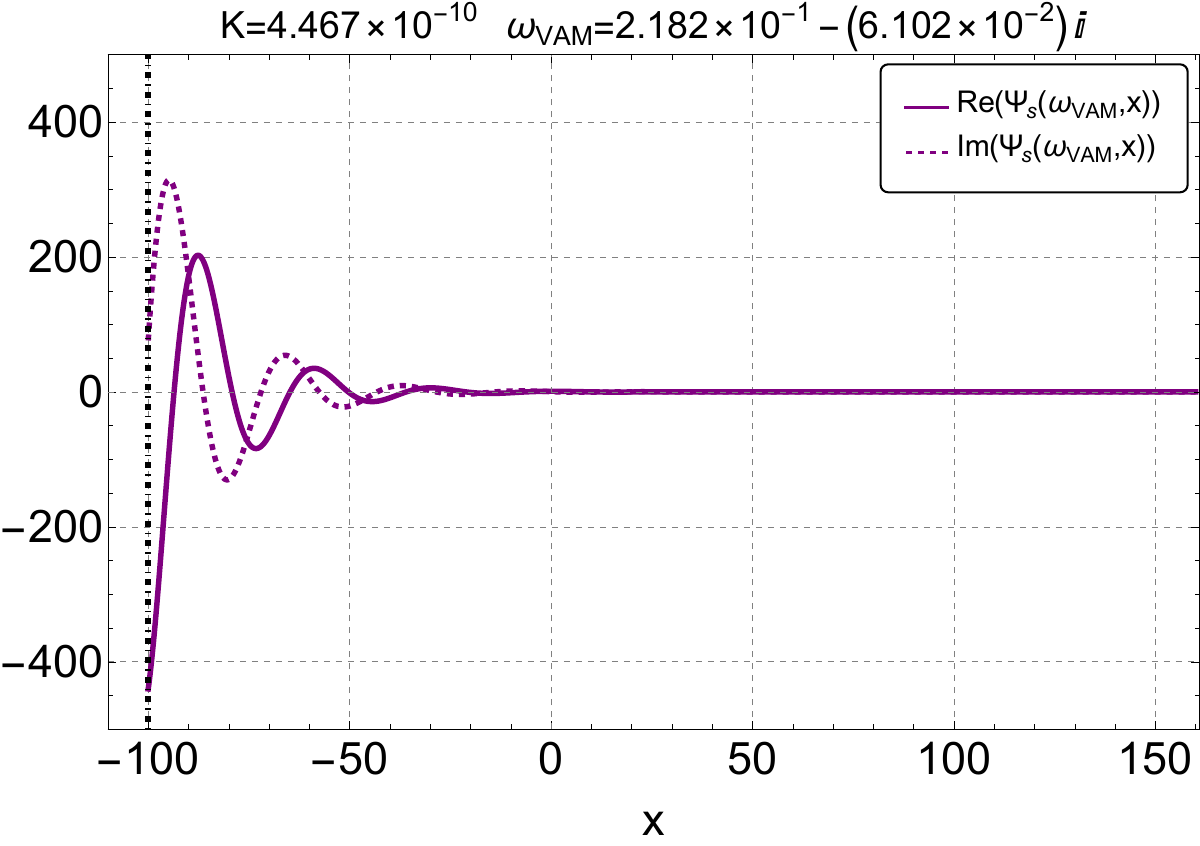}}
    \subfigure[]{\includegraphics[width=0.31\linewidth]{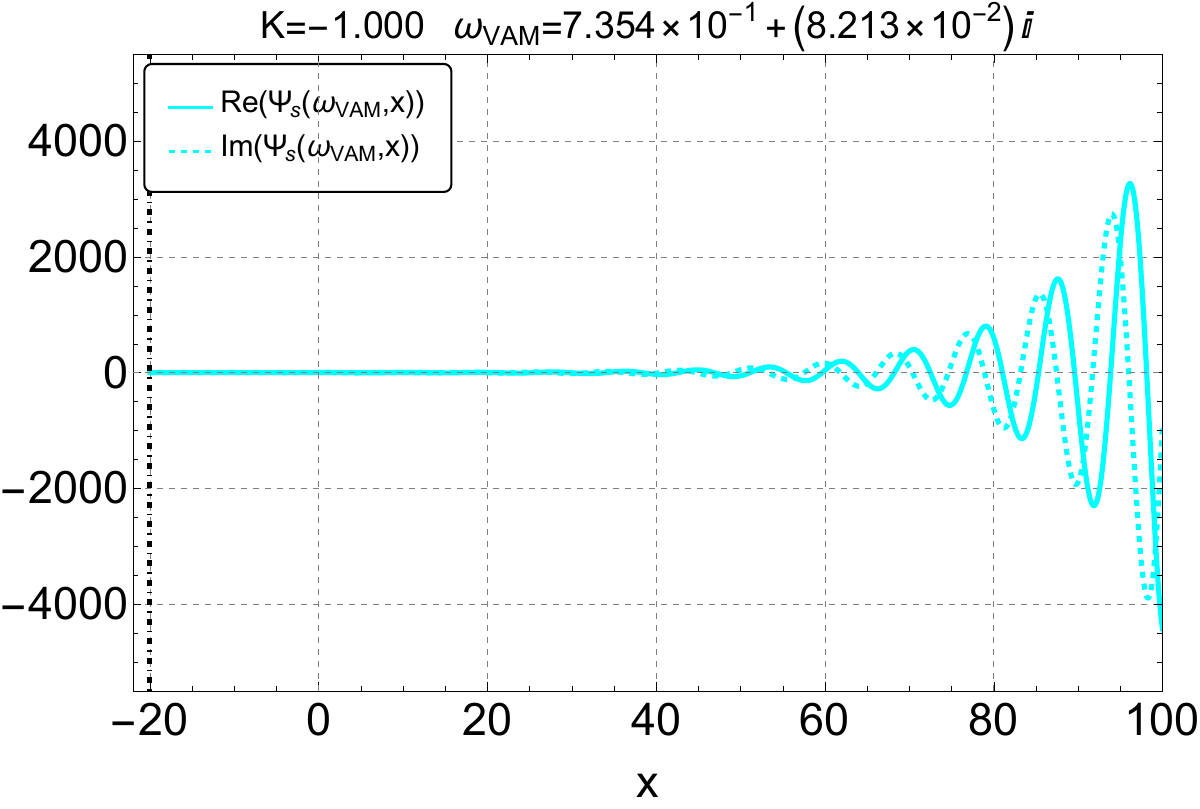}}\hfill
    \subfigure[]{\includegraphics[width=0.31\linewidth]{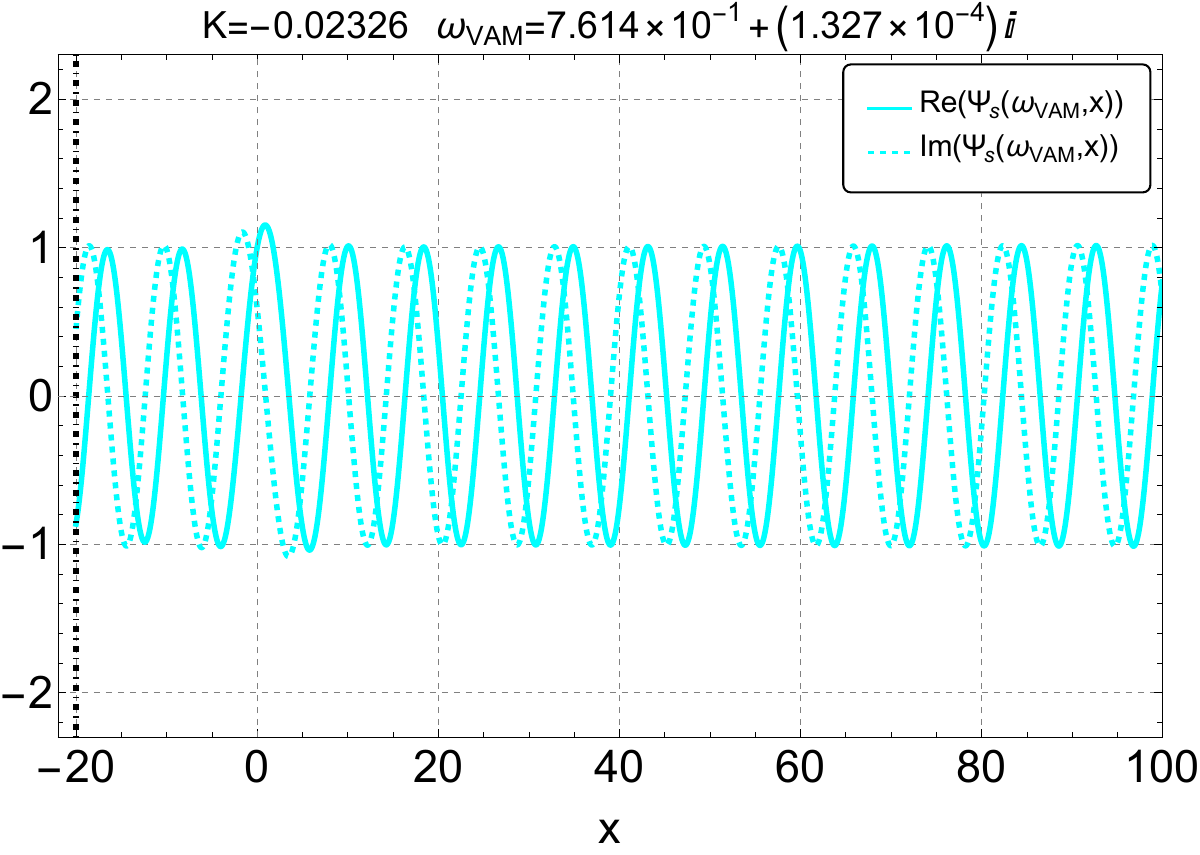}}\hfill
    \subfigure[]{\includegraphics[width=0.31\linewidth]{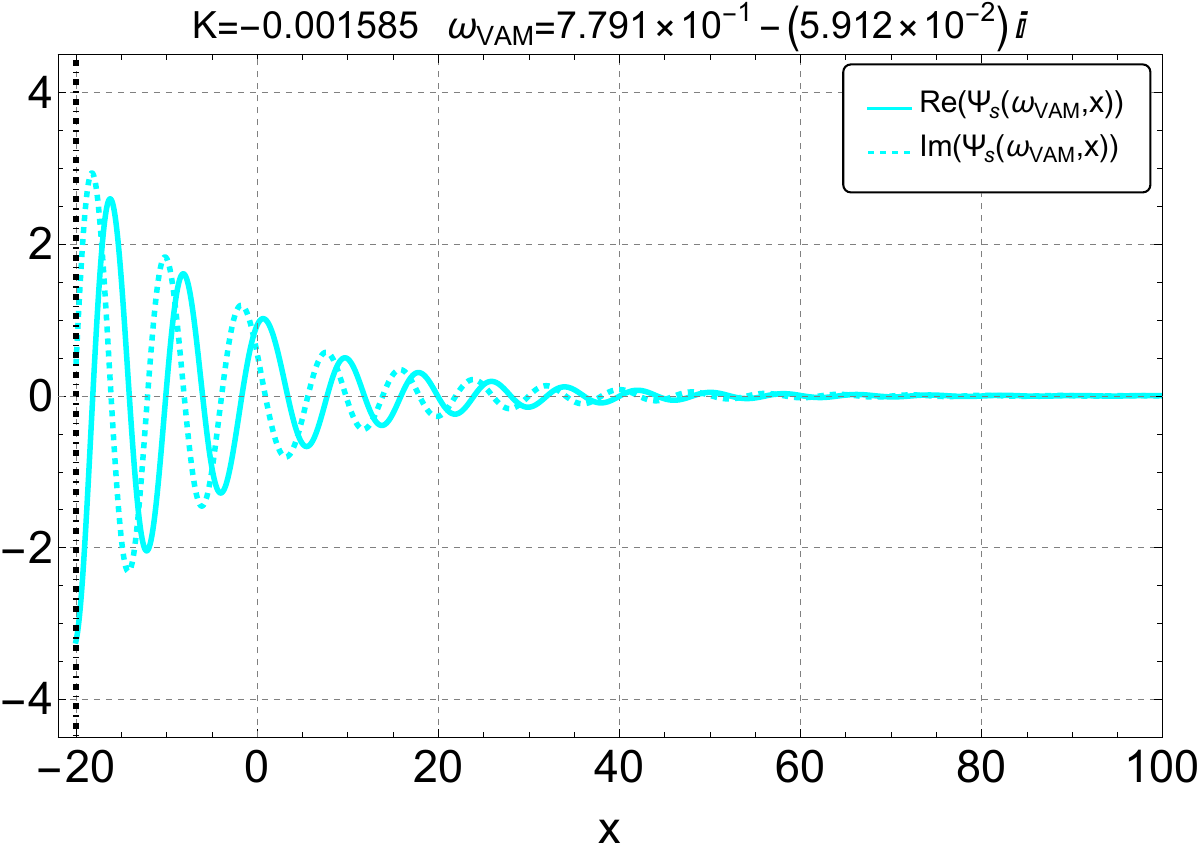}}
    \caption{The VAM eigenfunctions are shown. For the three top panels, we choose the modes $n=6$ and parameters being $r_\text{c}=1.1$, $x_{\text{w}}=-100$ with $\mathcal{K}>0$. For the three bottom panels, we choose the modes $n=4$ and parameters being $r_\text{c}=2$, $x_{\text{w}}=-20$ with $\mathcal{K}<0$. The solid line represents the real part of the eigenfunction, while the dashed line represents the imaginary part of the eigenfunction. The black thick dotted line is used to depict the position of the reflective wall.}
    \label{VAM_functions}
\end{figure}

As $\omega=\omega_{\text{VAM}}$, from $S_{\text{out}}(\omega_{\text{VAM}})=0$, it can be seen that the solution $\Psi_s(x)$ is proportional to the ``down'' solution $\Psi_{\text{down}}(x)$ by comparing Eq. (\ref{down_solution}) and Eq. (\ref{semi_open_conditions}). The proportionality coefficient will be $S_{\text{in}}(\omega_{\text{VAM}})$, i.e., 
\begin{eqnarray}
    \Psi_{s}(\omega_{\text{VAM}},x)=S_{\text{in}}(\omega_{\text{VAM}})\Psi_{\text{down}}(\omega_{\text{VAM}},x)=S_{\text{in}}(\omega_{\text{VAM}})Y_{\text{down}}(\omega_{\text{VAM}})\Psi_{12}(\omega_{\text{VAM}},x)\, ,\quad x\in\mathbb{R}\, ,
\end{eqnarray}
where the function $Y_{\text{down}}(\omega)$ in terms of $\omega$ is given by
\begin{eqnarray}\label{Ydown}
    Y_{\text{down}}(\omega)=(-1)^{\rho_{\text{c},2}(\omega)}(-1)^{\rho_{n,2}(\omega)}\times\Big(1-\frac{r_\text{e}}{r_{\text{c}}}\Big)^{\rho_{\text{e},2}(\omega)}\Big(\frac{r_{\text{c}}}{r_{\text{e}}}-1\Big)^{\rho_{\text{c},2}(\omega)}\Big(\frac{r_n}{r_\text{c}}-\frac{r_n}{r_\text{e}}\Big)^{\rho_{n,2}(\omega)}\Big[\frac{r_\text{e}(r_n-r_\text{c})}{(r_\text{c}-r_\text{e})r_n}\Big]^{1-\epsilon(\omega)}\, .
\end{eqnarray}
One can find the expressions of $S_{\text{in}}(\omega)$, $\rho_{\text{e},2}(\omega)$, $\rho_{\text{c},2}(\omega)$, $\rho_{n,2}(\omega)$ and $\epsilon(\omega)$ in~\cite{Wu:2025wbp}. In addition, the function $\Psi_{12}(x)$ consists of Heun's function~\cite{Wu:2025wbp}. Since the sign of the imaginary part of $\omega_{\text{VAM}}$ changes with $\mathcal{K}$, which will also influence the asymptotic behaviors ($x\to+\infty$ and $x\to x_{\text{w}}$) of the eigenfunctions, we display  behaviors of the eigenfunctions corresponding to cases where the imaginary part of $\omega_{\text{VAM}}$ is greater than zero, close to zero, and less than zero. In Fig. \ref{VAM_functions}, several VAM eigenfunctions $\Psi_{s}(\omega_{\text{VAM}},x)$ are depicted, in which three top panels stand for the results of parameters with $r_{\text{c}}=1.1$, $x_{\text{w}}=-100$ and $n=6$ (purple color), while three bottom panels stand for the results of parameters with $r_{\text{c}}=2$, $x_{\text{w}}=-20$ and $n=4$ (cyan color). 

\section{Absorption and release of energy}\label{sec: simulation}
In this section, we will study the absorption and release of energy in the semi-open Schwarzschild-de Sitter spacetime~\cite{Tuncer:2025dnp}. By meticulously tailoring the initial conditions of an incoming wave, rendering it non-monochromatic in the conventional sense (real frequency), virtual absorption phenomenon can be achieved. In this process, the incident waveform is entirely absorbed by the medium, with no reflection. Virtual absorption phenomena have been extensively studied in various physical systems, including optical~\cite{Baranov:2017}, acoustic~\cite{Maddi:2025}, elastodynamic~\cite{Trainiti:2019}, and quantum~\cite{Zhang:2026jsc}. Recently, it is also been extended to gravitational systems through the concepts of virtual absorption modes (VAMs) and total absorption modes (TAMs) introduced in~\cite{Tuncer:2025dnp}.  

This requires us to conduct a time-domain analysis, i.e., solving Eq. (\ref{master_equation_time_domain}) to get the waveform under reflective boundary conditions (\ref{semi_open_conditions}). Combining with the time convention $\mathrm{e}^{-\mathrm{i}\omega t}$, one has the boundary condition at the wall been written as
\begin{eqnarray}\label{boundary_condition_wall}
    \frac{\mathrm{\partial}\Psi}{\mathrm{\partial }x}\Bigg|_{x_\text{w}}=\mu\frac{\partial \Psi}{\partial t}\Bigg|_{x_\text{w}}\, ,\quad \mu\equiv \frac{1-\mathcal{K}}{1+\mathcal{K}}\, ,
\end{eqnarray}
for the constant reflectivity $\mathcal{K}\neq-1$. The outgoing boundary condition at the infinity (cosmological horizon) is written as
\begin{eqnarray}\label{boudary_condition_infinity}
    \lim_{x\to+\infty}\Big(\frac{\partial}{\partial t}+\frac{\partial}{\partial x}\Big)\Psi(t,x)=0\, .
\end{eqnarray}
Note that as $\mathcal{K}=-1$, we have the (homogeneous) Dirichlet boundary condition, namely $\Psi(t,x_\text{w})=0$. Eq. (\ref{master_equation_time_domain}) with boundary conditions (\ref{boundary_condition_wall}) and (\ref{boudary_condition_infinity}) can be solved by the method of line (MOL), which is described in Appendix \ref{numerical_method_TD}. In the numerical implementation, we place the outer boundary at a sufficiently large value of $x_{\text{max}}$ to approximate spatial infinity.

For Eq. (\ref{master_equation_time_domain}), it is not difficult to find that the corresponding energy density $\rho(t,x)$ is
\begin{eqnarray}\label{energy_density}
    \rho(t,x)=\frac{1}{2}\Big[|\partial_t\Psi|^2+|\partial_x\Psi|^2+V_s(x)|\Psi|^2\Big]=\frac{1}{2}\Big[|\Pi|^2+|\partial_x\Psi|^2+V_s(x)|\Psi|^2\Big]\, ,
\end{eqnarray}
and the energy flux density $j(t,x)$ is 
\begin{eqnarray}\label{energy_flux_density}
    j(t,x)=-\frac{1}{2}\Big(\partial_t\Psi^{\star}\cdot\partial_x\Psi+\partial_t\Psi\cdot\partial_x\Psi^{\star}\Big)=-\frac{1}{2}\Big(\Pi^{\star}\cdot\partial_x\Psi+\Pi\cdot\partial_x\Psi^{\star}\Big)\, ,
\end{eqnarray}
in which $\star$ represents the complex conjugate and we have defined $\Pi(t,x)\equiv\partial_t\Psi(t,x)$. At a fixed time $t$, the total energy on the considered spatial interval $[x_{\text{w}},x_{\text{max}}]$ is the integration over this interval
\begin{eqnarray}\label{energy}
    E_{\text{total}}(t)=\int_{x_{\text{w}}}^{x_{\text{max}}}\rho(t,x)\mathrm{d}x=E_{\text{inside}}(t)+E_{\text{outside}}(t)\, ,
\end{eqnarray}
where
\begin{eqnarray}\label{energy_in_out}
    \quad E_{\text{inside}}(t)=\int_{x_{\text{w}}}^{x_{\text{cut}}}\rho(t,x)\mathrm{d}x\, ,\quad  E_{\text{outside}}(t)=\int_{x_{\text{cut}}}^{x_{\text{max}}}\rho(t,x)\mathrm{d}x\, .
\end{eqnarray}
An effective potential and reflective wall together constitute an energy absorber, with the amount of stored energy characterized by $E_{\text{inside}}(t)$. The peak of the effective potential serves as the outer wall of the cavity formed between the reflective boundary and the potential barrier. Energy is stored inside this cavity during virtual absorption. Therefore, $x_{\text{cut}}$ is chosen as the extreme point of the effective potential. The dynamic changes of $E_{\text{total}}(t)$, $E_{\text{inside}}(t)$ and $E_{\text{outside}}(t)$ over time reflect the process of energy absorption and release. From the conservation of energy, namely $\partial_t\rho+\partial_xj=0$, one has
\begin{eqnarray}\label{energy_derivative_t}
    \frac{\mathrm{d}E_{\text{total}}(t)}{\mathrm{d}t}=-\mu\Big|\Pi(t,x_{\text{w}})\Big|^2-\Big|\partial_x\Psi(t,x_{\text{max}})\Big|^2\, ,
\end{eqnarray}
where the boundary conditions (\ref{boundary_condition_wall}) and (\ref{boudary_condition_infinity}) have been used. Note that the boundary condition for QNMs corresponds to $\mu=1$. This is consistent with the widely recognized notion that QNM systems are dissipative. Because for $\mu>0$ (i.e., $|\mathcal{K}|<1$), we have $\mathrm{d}E_{\text{total}}(t)/\mathrm{d}t\leq0$. Energy is dissipated, and the boundary acts as an impedance absorber. 

Suppose that the local waveform always moves away from the outer boundary during propagation, then we have $\partial_x\Psi(t,x_{\text{max}})\approx0$. For $\mu<0$ (i.e., $|\mathcal{K}|>1$), $\mathrm{d}E_{\text{total}}(t)/\mathrm{d}t\geq0$. Energy increases, meaning the inner boundary injects energy into the system. For $\mu=0$ (i.e., $\mathcal{K}=1$), which corresponds to the homogeneous Neumann condition, the energy is conserved: $\mathrm{d}E_{\text{total}}(t)/\mathrm{d}t=0$. Note that for the Dirichlet boundary condition $\Psi(t,x_\text{w})=0$, one also has $\Pi(t,x_{\text{w}})=0$, which means that the energy is also conserved. If we consider an inhomogeneous Dirichlet boundary condition (for example $\Psi(t,x_{\text{w}})=g(t)\neq0$), then the energy flux at the inner boundary is generally non-zero, and its specific form depends on $g(t)$ and the solution to the wave equation (\ref{master_equation_time_domain}). In this case, the inner boundary may inject energy into or absorb energy from the system.

In previous section, VAM spectra have been solved. Now, we excite the system with a specific VAM with the frequency $\Omega_0$. To achieve this, we consider the following family of initial data, which consists of an exponentially growing sinusoidal wave truncated by a Gaussian envelope at a truncated radius $x_{\text{t}}$~\cite{Tuncer:2025dnp}, 
\begin{eqnarray}
    \Psi(0,x)=\mathcal{A}\Big[\mathrm{e}^{\omega_{\text{I}}(x-x_{\text{t}})}\Theta(x_{\text{t}}-x)+\mathrm{e}^{-(x-x_{\text{t}})^2/(2\sigma^2)}\Theta(x-x_{\text{t}})\Big]\cos\Big[\omega_{\text{R}}(x-x_{\text{t}})\Big]\, ,\quad \Pi(0,x)=\frac{\mathrm{d}\Psi(0,x)}{\mathrm{d}x}\, ,
\end{eqnarray}
where $\Omega_0=\omega_{\text{R}}+\mathrm{i}\omega_{\text{I}}$ with $\omega_{\text{I}}>0$, $\Theta$ is the Heaviside step function, and the maximum of $\Psi(0,x)$, $\mathcal{A}$, is referred to as the amplitude. The selection of the initial condition for $\Pi$ is to ensure that the initial waveform propagates towards the effective potential to simulate energy injection. Note that since initial data is real, all the numerical data obtained will be real.

\begin{figure}[htbp]
    \centering
    \subfigure[]{\includegraphics[width=0.45\linewidth]{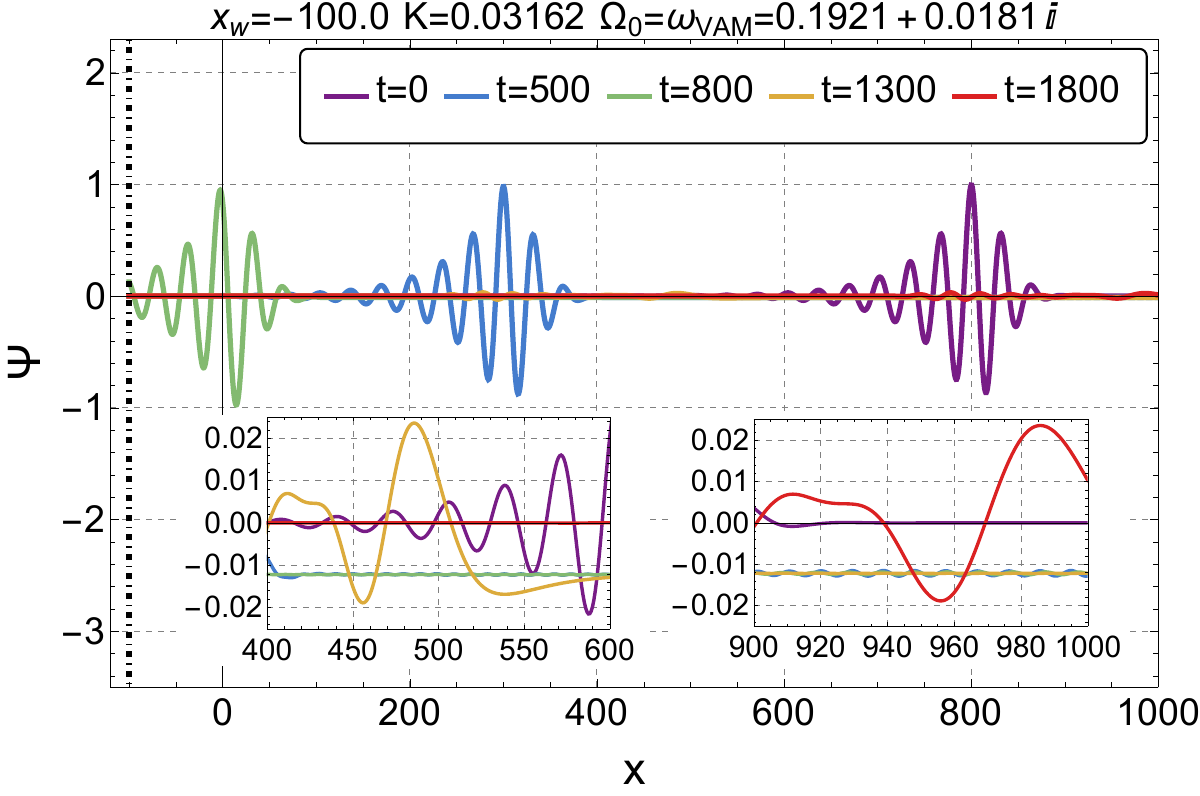}\label{Energy_Plot_rc_1p1_a}}\hfill
    \subfigure[]{\includegraphics[width=0.45\linewidth]{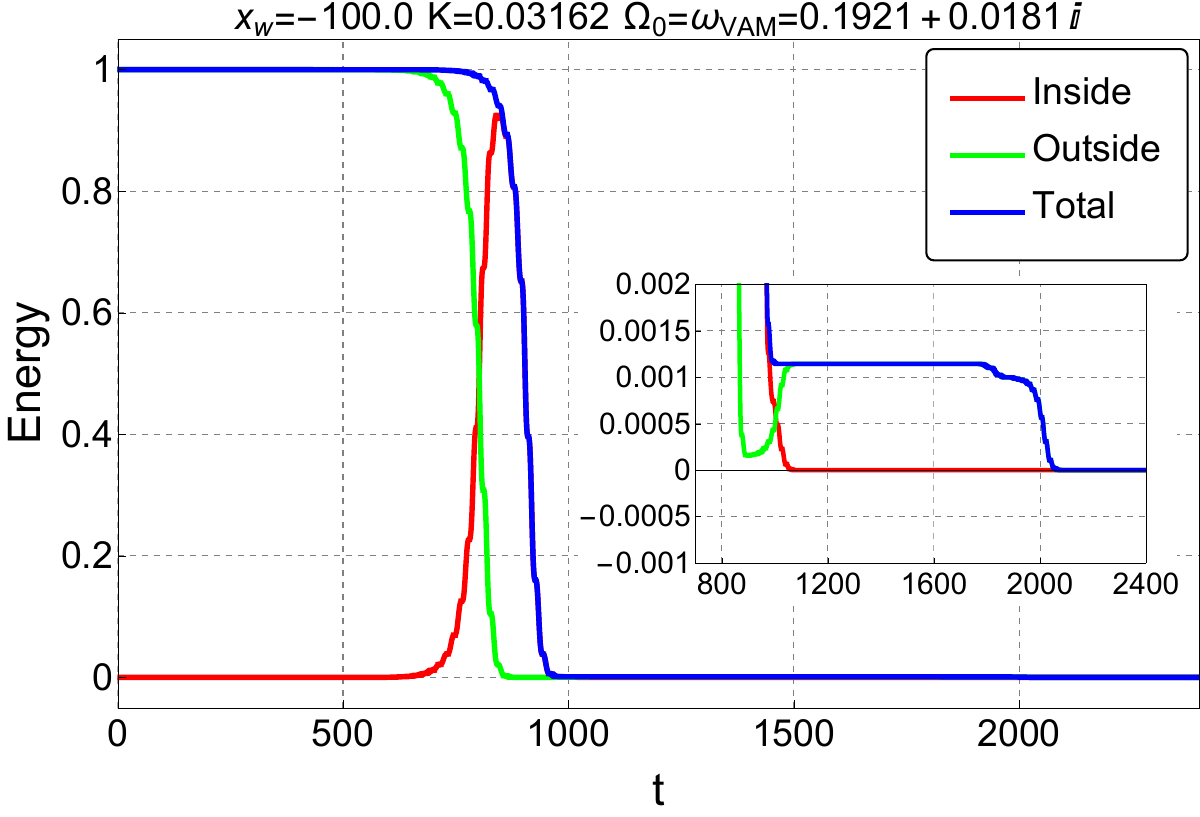}\label{Energy_Plot_rc_1p1_b}}
    \subfigure[]{\includegraphics[width=0.45\linewidth]{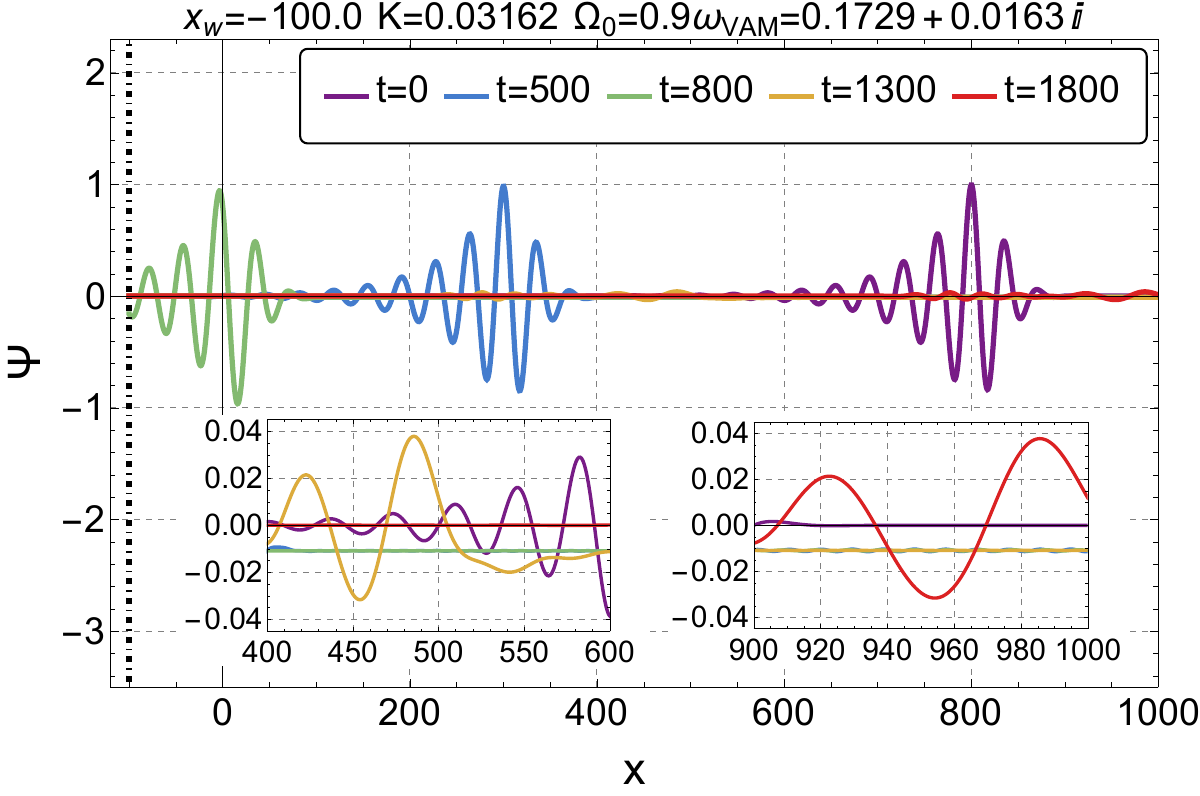}\label{Energy_Plot_rc_1p1_c}}\hfill
    \subfigure[]{\includegraphics[width=0.45\linewidth]{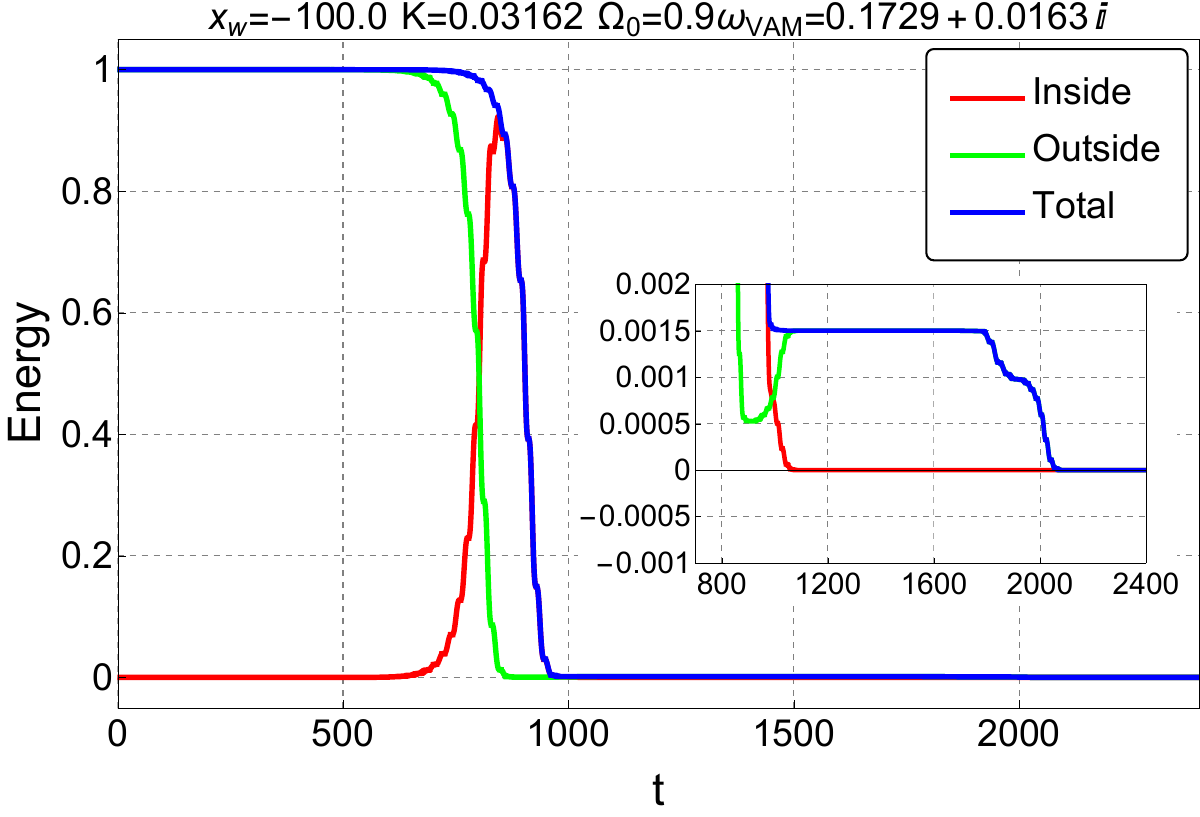}\label{Energy_Plot_rc_1p1_d}}
    \subfigure[]{\includegraphics[width=0.45\linewidth]{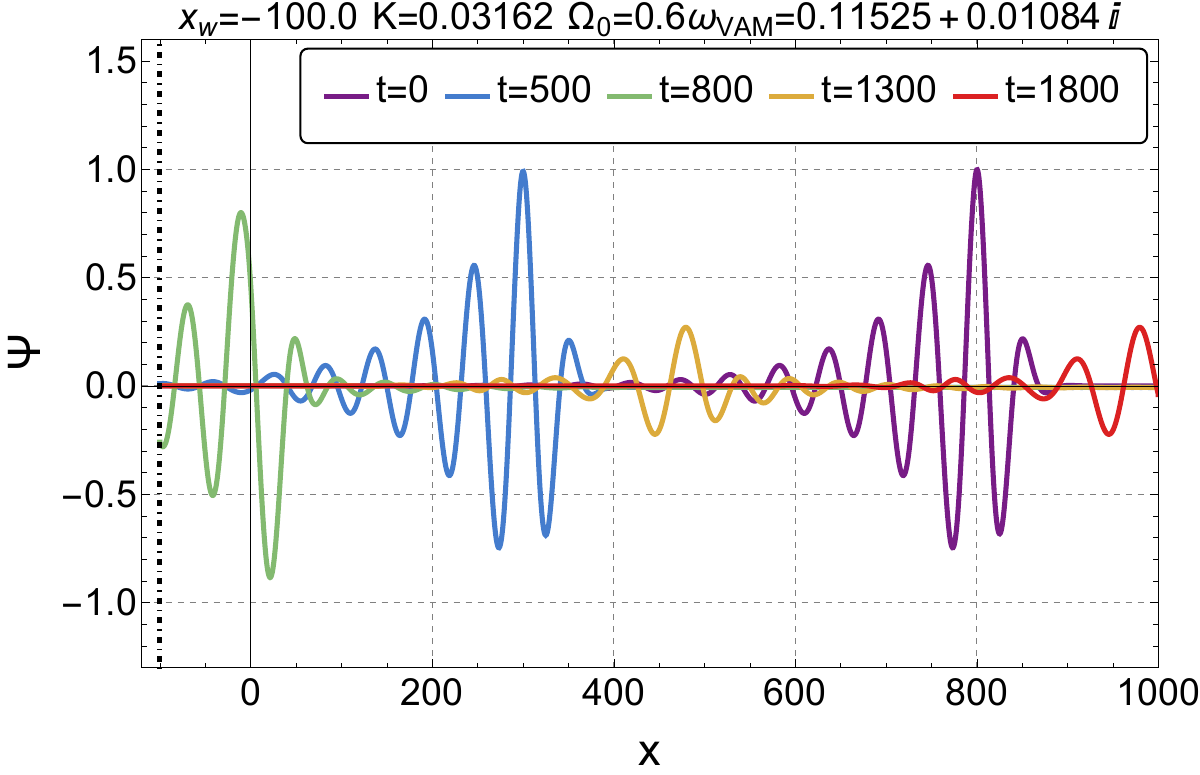}\label{Energy_Plot_rc_1p1_e}}\hfill
    \subfigure[]{\includegraphics[width=0.45\linewidth]{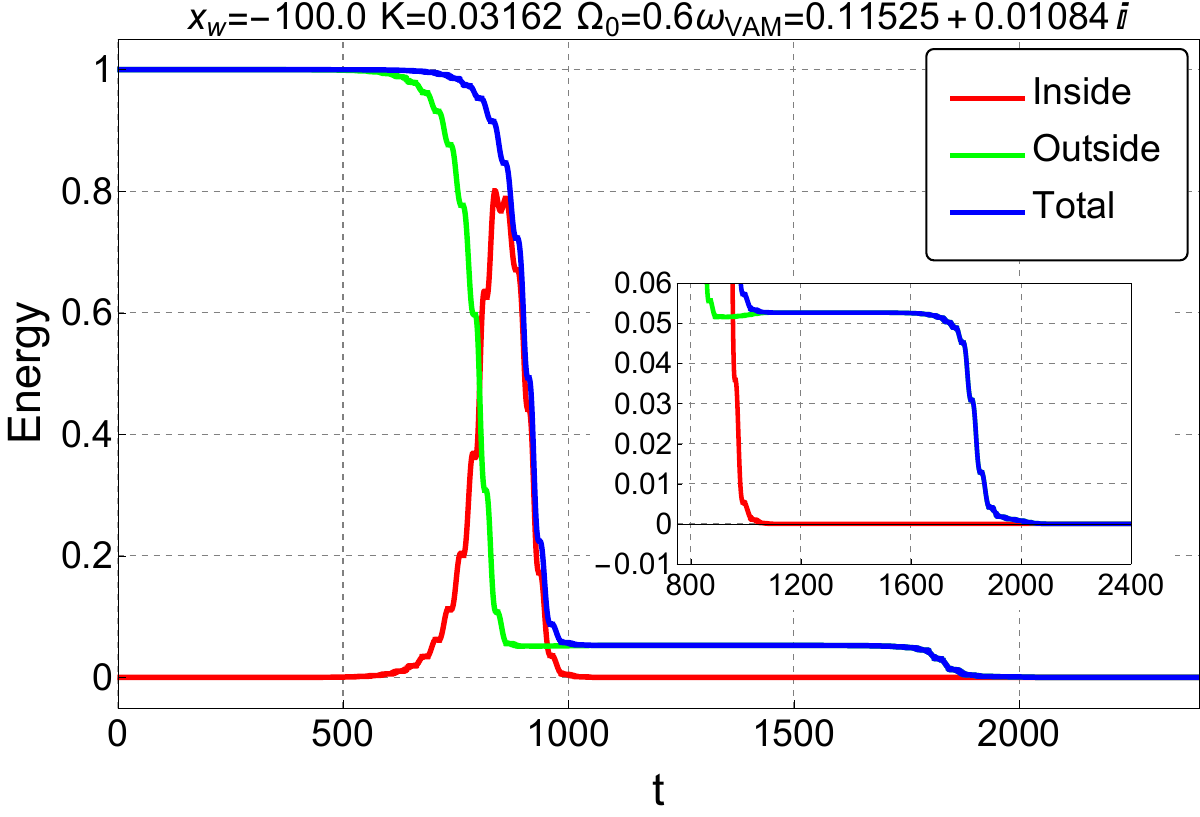}\label{Energy_Plot_rc_1p1_f}}
    \caption{The time evolution of $\Psi(t,x)$ and energy are shown, where the parameters are $r_\text{c}=1.1$, $x_{\text{w}}=-100$ and $\mathcal{K}=0.03162$. The parameters of the initial conditions are $\mathcal{A}=1$, $x_{\text{t}}=800$ and $\sigma=30$. In the top panels, the overtone $n=6$ mode $\omega_{\text{VAM}}=0.1921+0.0181\mathrm{i}$ is excited (purple solid line in Fig. \ref{TTM_Spectra_rc_1p1_a}), while in the middle and bottom panels, non-VAM spectrum with $\Omega_0=0.9\omega_{\text{VAM}}=0.1729+0.0163\mathrm{i}$ and non-VAM spectrum with $\Omega_0=0.6\omega_{\text{VAM}}=0.11525+0.01084\mathrm{i}$ are excited for comparison. In addition, numerical parameters are $\mathrm{d}t=0.1$ and $N=350$, and $x_{\text{cut}}=1.76$ is used to distinguish between the inside and outside aspects of effective potential. The black thick dotted line is used to depict the position of the reflective wall.}
    \label{Energy_Plot_rc_1p1}
\end{figure}

\begin{figure}[htbp]
    \centering
    \subfigure[]{\includegraphics[width=0.45\linewidth]{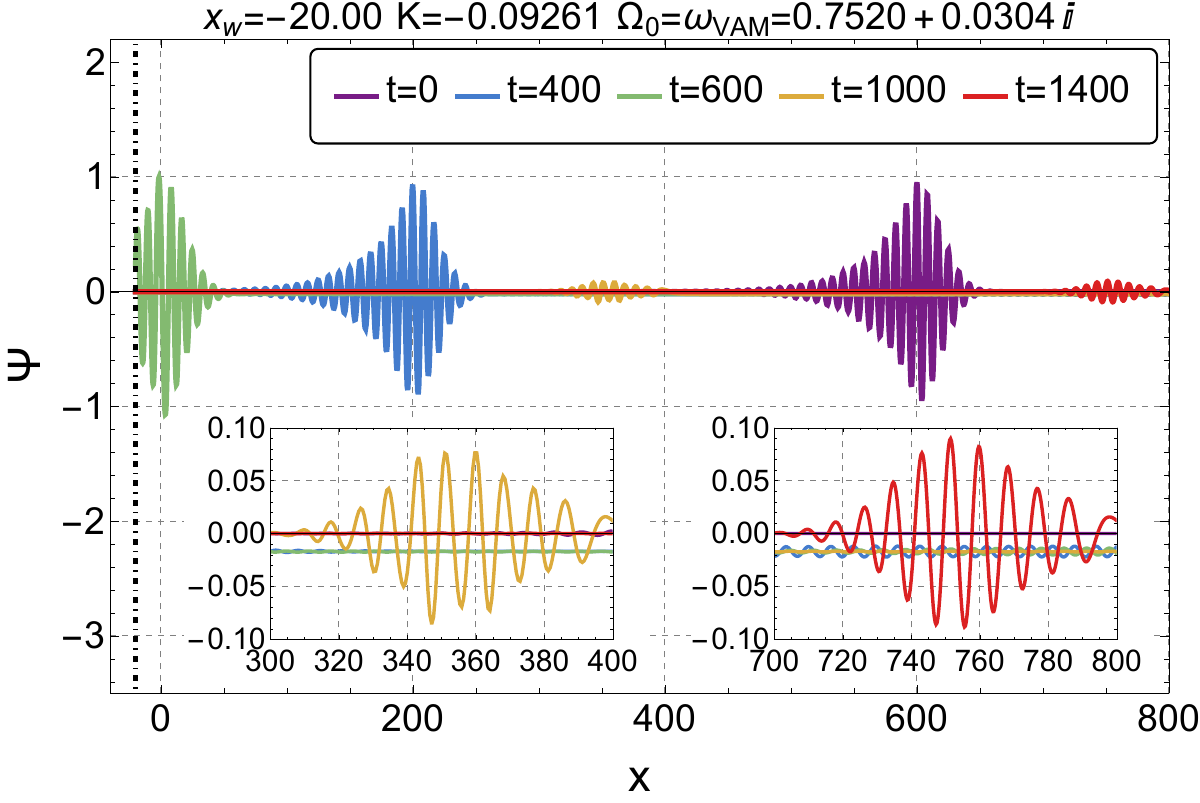}}\hfill
    \subfigure[]{\includegraphics[width=0.45\linewidth]{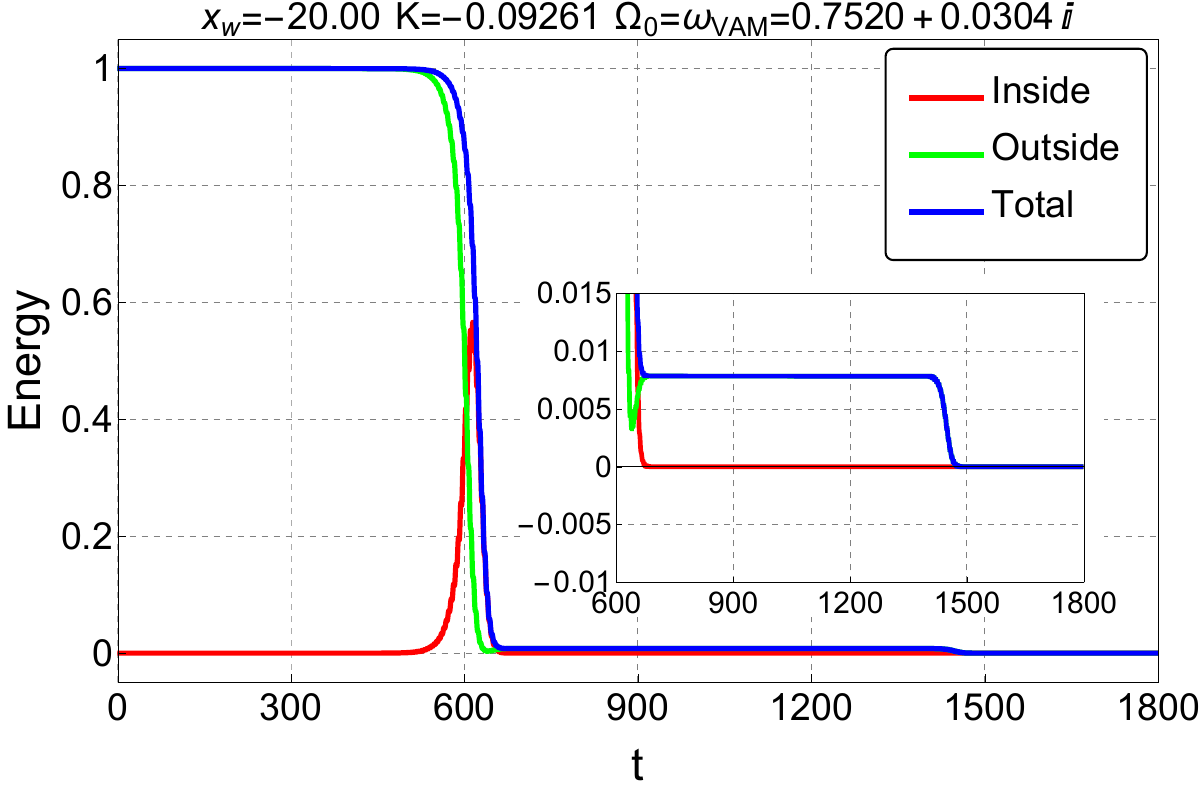}}
    \subfigure[]{\includegraphics[width=0.45\linewidth]{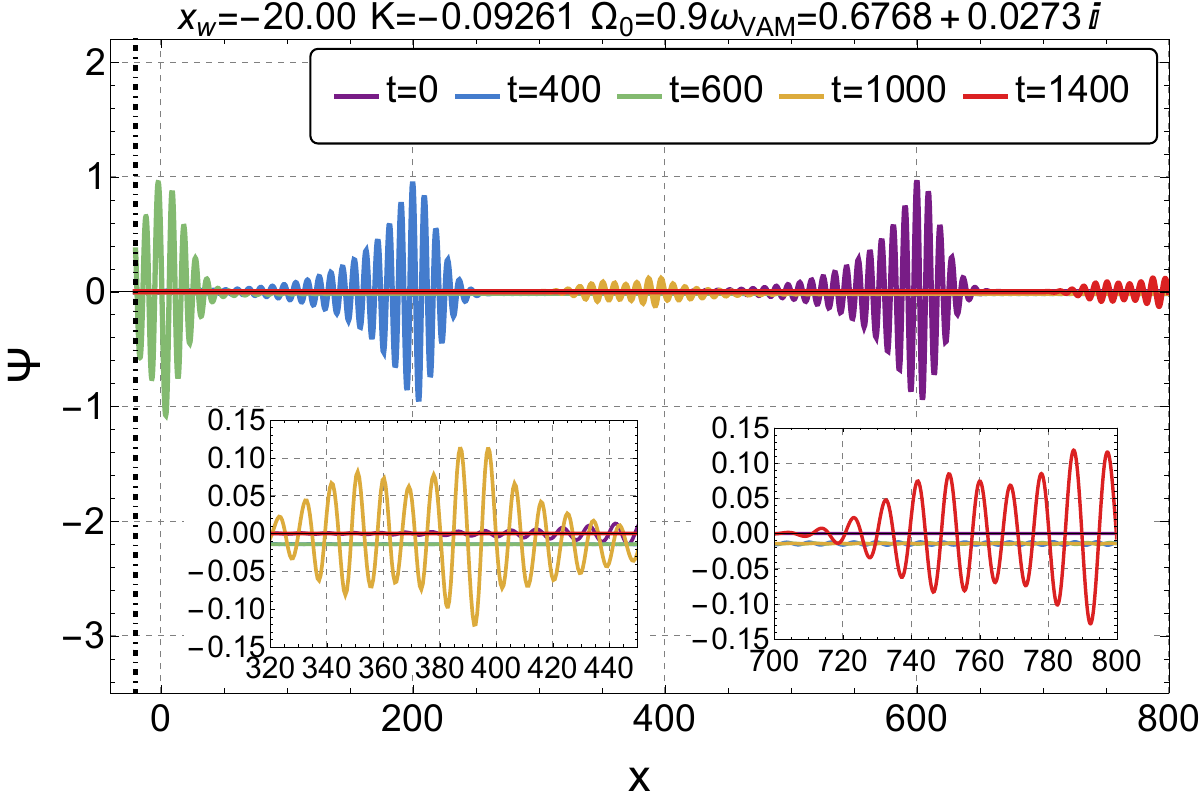}}\hfill
    \subfigure[]{\includegraphics[width=0.45\linewidth]{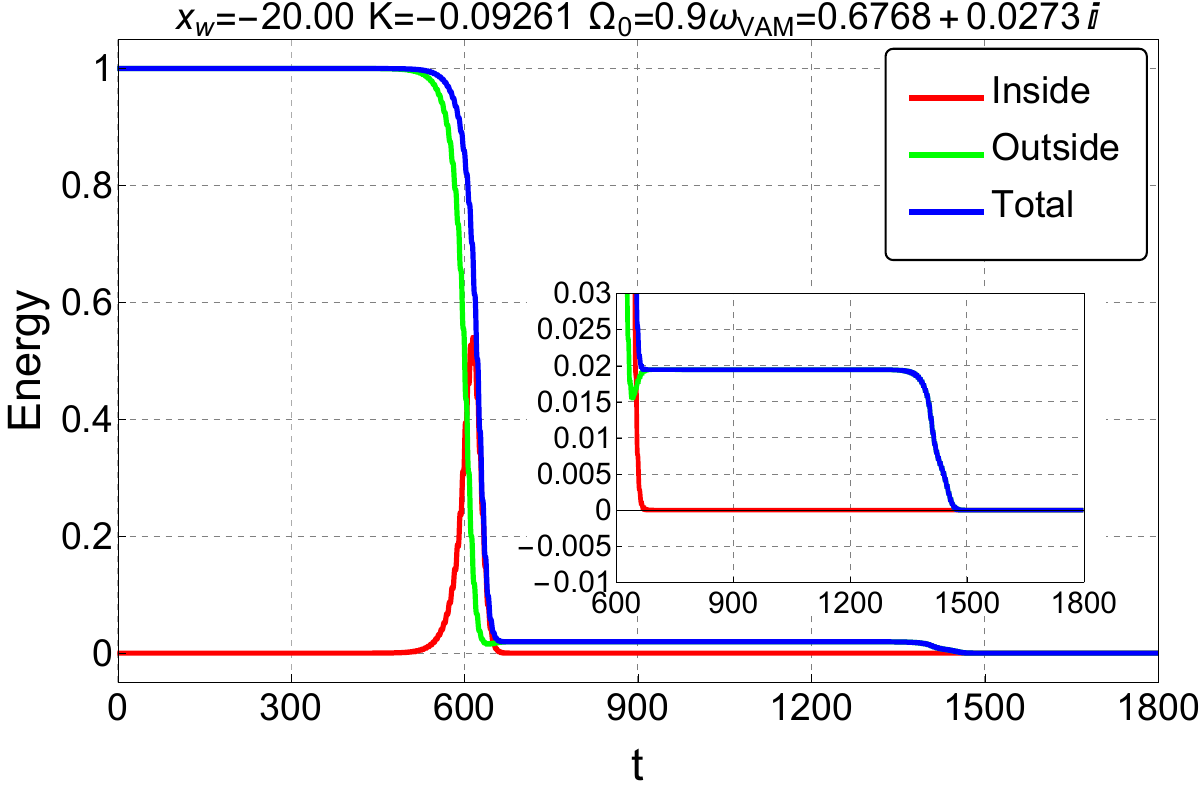}}
    \subfigure[]{\includegraphics[width=0.45\linewidth]{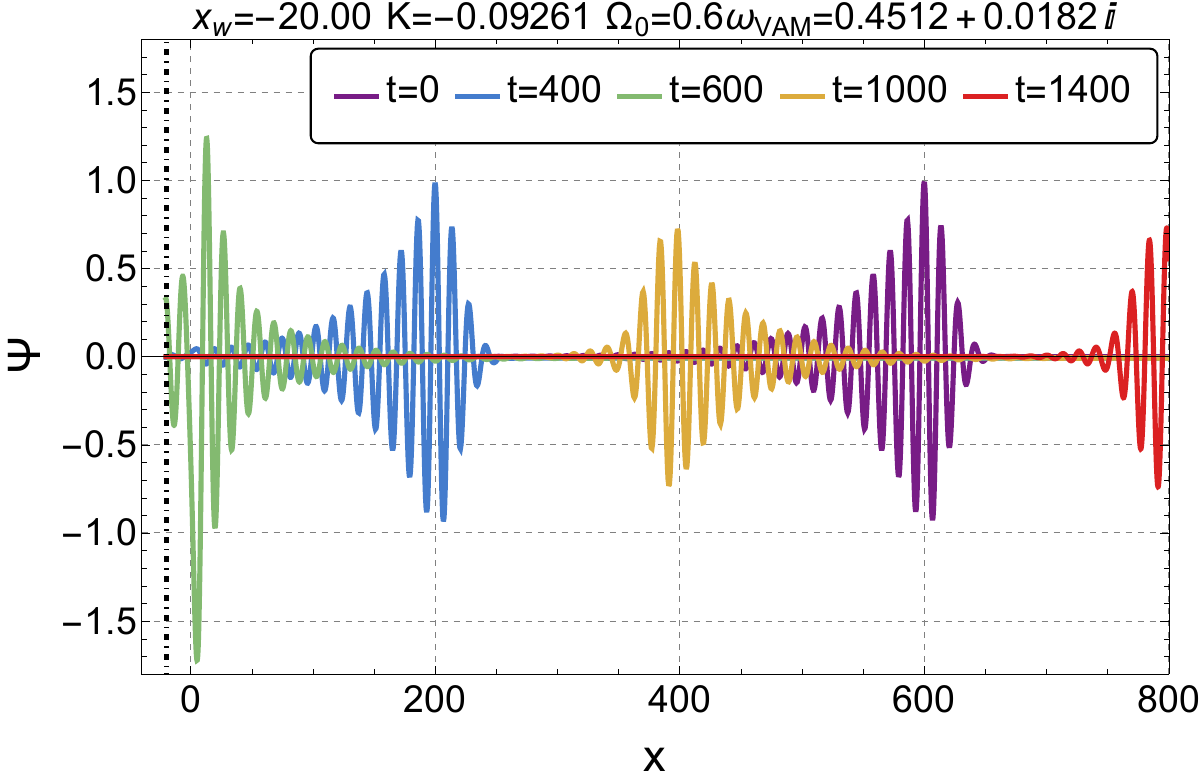}}\hfill
    \subfigure[]{\includegraphics[width=0.45\linewidth]{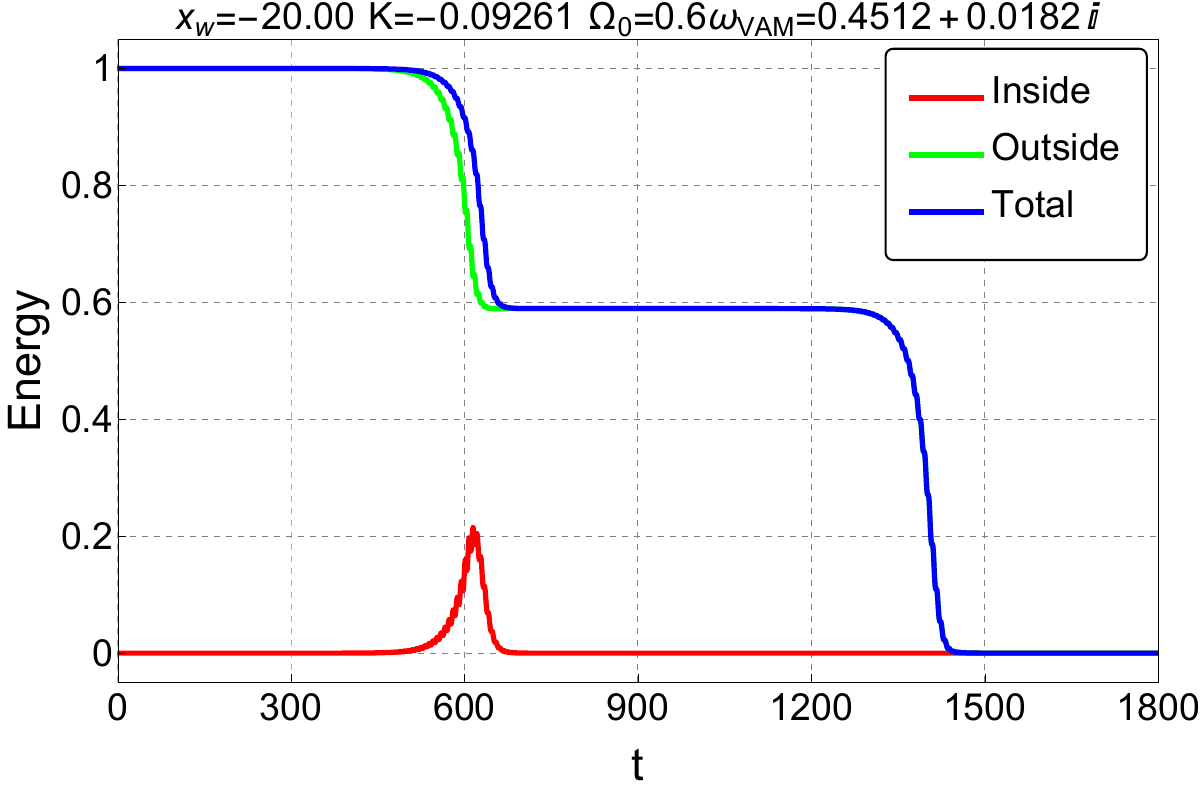}}
    \caption{The time evolution of $\Psi(t,x)$ and energy are shown, where the parameters are $r_\text{c}=2$, $x_{\text{w}}=-20$ and $\mathcal{K}=-0.09261$. The parameters of the initial conditions are $\mathcal{A}=1$, $x_{\text{t}}=600$ and $\sigma=18$. In the top panels, the overtone $n=4$ mode $\omega_{\text{VAM}}=0.7520+0.0304\mathrm{i}$ is excited (cyan dashed line in Fig. \ref{TTM_Spectra_rc_2_a}), while in the middle and bottom panels, non-VAM spectrum with $\Omega_0=0.9\omega_{\text{VAM}}=0.6768+0.0273\mathrm{i}$ and non-VAM spectrum with $\Omega_0=0.6\omega_{\text{VAM}}=0.4512+0.0182\mathrm{i}$ are excited for comparison. In addition, numerical parameters are $\mathrm{d}t=0.1$ and $N=390$, and $x_{\text{cut}}=0.602$ is used to distinguish between the inside and outside aspects of effective potential. It is worth mentioning that in this simulation, the real part of mode is larger, so a higher resolution is required. The black thick dotted line is used to depict the position of the reflective wall.}
    \label{Energy_Plot_rc_2}
\end{figure}

The absorber, formed by the effective potential and the reflective wall, begins to absorb incoming energy as soon as the initial waveform tailored to excite the VAM reaches the effective potential, and continues until the excitation ends. Once the incident wave is turned off, the stored energy will be released back into space in the form of exponential decay sinusoidal wave. In other words, with infinite excitation time, there would be no reflection from the absorber. Fig. \ref{Energy_Plot_rc_1p1} and Fig. \ref{Energy_Plot_rc_2} illustrate the time evolution of the waveform $\Psi(t,x)$ and the corresponding energy evolution in a semi-open SdS spacetime under the excitation of VAMs. For the convenience of comparison, we have normalized the energy to $1$ at the initial moment. These figures provide a visual representation of the absorption and release of energy, demonstrating the virtual absorption phenomenon associated with VAM excitation in a curved spacetime with a cosmological constant. 

In Fig. \ref{Energy_Plot_rc_1p1}, the parameters are set to cosmological horizon $r_\text{c}=1.1$, reflective wall position $x_{\text{w}}=-100$, and reflectivity $\mathcal{K}=0.03162$. The top panels [Fig. \ref{Energy_Plot_rc_1p1_a} and Fig. \ref{Energy_Plot_rc_1p1_b}] show the dynamic system excited precisely at the VAM spectrum, and the middle and bottom panels [Fig. \ref{Energy_Plot_rc_1p1_c}-Fig. \ref{Energy_Plot_rc_1p1_f}] show the dynamic system excited at the non-VAM spectra. The spatial profiles of $\Psi(t,x)$ at various time reveals clear patterns. We can see that the waveform maintains a relatively stable shape and moves to the left before hitting the effective potential. After the waveform hits the effective potential, the absorber begins to absorb energy. For the total energy $E_{\text{total}}$, since $|\mathcal{K}|$ is not equal to $1$, it always decreases (cf. Eq. (\ref{energy_derivative_t})). The decrease in total energy can also be divided into two stages. The first stage is due to the absorption effect of the reflective boundary, and the second stage is due to the waveform leaving the computational domain. For the outside energy $E_{\text{outside}}(t)$, it first decreases to a local minimum denoted by $\eta$, then increases to a certain value and finally decays to $0$. The reason for the decrease of $E_{\text{outside}}(t)$ to a local minimum is also due to the absorber continuously absorbing energy. The intermediate platform of $E_{\text{outside}}(t)$ corresponds to the energy of the reflected waveform. However, most importantly, it is found that when $\Omega_0$ is the VAM spectrum, the local minimum of outside energy is minimized [see Fig. \ref{Energy_Plot_rc_1p1_b}]. Conversely, when $\Omega_0$ is not the VAM spectrum, the local minimum of outside energy is relatively larger [see Fig. \ref{Energy_Plot_rc_1p1_d} and Fig. \ref{Energy_Plot_rc_1p1_f}]. Quantitatively speaking, the values of $\eta$ in Fig. \ref{Energy_Plot_rc_1p1_b}, Fig. \ref{Energy_Plot_rc_1p1_d} and Fig. \ref{Energy_Plot_rc_1p1_f} are approximately $1.564\times10^{-4}$, $5.254\times10^{-4}$, and $5.158\times10^{-2}$, respectively\footnote{For this setting, we have tested other $\Omega_0$'s, although the figures of the results are not shown. For $\Omega_0=\omega_{\text{VAM}}-0.01\mathrm{i}$, we have $\eta=3.116\times10^{-4}$. For $\Omega_0=\omega_{\text{VAM}}-0.05$, we have $\eta=6.901\times10^{-3}$.}. Furthermore, the resolution tests are presented in Appendix \ref{Numerical_convergence_test}. The virtual absorption effect becomes most significant when the system is excited precisely at the VAM spectrum.

To illustrate the specificity of the phenomenon of virtual absorption, we present the results of another set of parameters, namely Fig. \ref{Energy_Plot_rc_2}. For Fig. \ref{Energy_Plot_rc_2}, parameters are set to cosmological horizon $r_c=2$, reflective wall position $x_{\text{w}}=-20$, and reflectivity $\mathcal{K}=-0.09261$. The qualitative phenomenon is similar with Fig. \ref{Energy_Plot_rc_1p1}. Similarly, the corresponding values of $\eta$ in the top, middle, and bottom panels are $3.290\times10^{-3}$, $1.551\times10^{-2}$, and $0.5888$, respectively.

\section{Conclusions and discussion}\label{conclusions}
In this work, we present a comprehensive investigation of virtual absorption modes (VAMs) within the framework of semi-open Schwarzschild-de Sitter (SdS) spacetimes. By rigorously solving the perturbation equations governing axial gravitational perturbations, it is shown that VAMs emerge as a natural and physically significant spectrum family when reflective boundary conditions are imposed near the event horizon. Unlike QNMs, which are defined by purely outgoing waves at cosmological horizon and purely ingoing waves at the event horizon, VAMs are characterized by the condition of purely ingoing waves at cosmological horizon.

Utilizing the exact solvability of the axial perturbation equation via Heun’s functions, we derive the analytic expressions for the relevant homogeneous solutions, namely, the ``in'', ``up'', and ``down'' solutions~\cite{Wu:2025wbp}. The VAMs are then obtained by imposing the condition $S_{\text{out}}(\omega)=0$. At this moment, the reflective solution $\Psi_s$ is proportional to the ``down'' solution $\Psi_{\text{down}}$. Our numerical analysis reveals that the VAM spectra exhibit predictable behavior as a function of the reflectivity (see Fig. \ref{TTM_Spectra_rc_1p1} and Fig. \ref{TTM_Spectra_rc_2}). Remarkably, for each overtone, there exists a critical value $\mathcal{K}$  at which the imaginary part of the mode vanishes, corresponding to a purely oscillatory solution. In this case, these modes are called total absorption modes (TAMs)~\cite{Tuncer:2025dnp}. Furthermore, we have analyzed the influence of the cosmological horizon radius and the wall position on the VAM spectra. While quantitative differences arise such as the spacing between adjacent modes decreasing as the wall approaches the horizon, the qualitative features remain robust across different parameter regimes. This universality underscores the fundamental nature of VAMs in semi-open black hole spacetimes. 

A key result of our study is the demonstration that VAMs serve as the spectral signatures of virtual absorption processes. By performing time-domain simulations using a high-order Hermite integration method combined with Chebyshev spectral collocation, we have shown that when the system is excited precisely at a VAM spectrum, the reflected wave amplitude is minimized. These results confirm that when driven at the corresponding VAM spectra, the semi-open SdS system behaves as a perfect absorber, establishing a direct connection between the system's spectrum analysis and its time-domain energy dynamics. 

From a broader perspective, this work contributes to the ongoing effort to understand the spectral properties of black holes beyond the traditional QNMs. The introduction of reflective boundary conditions, motivated by models of exotic compact objects or quantum gravity effects near the horizon, enriches the spectral structure and opens new avenues for probing the nature of compact objects through gravitational wave observations. The concept of VAMs, originally introduced in the context of optical and acoustic systems, is here extended to SdS black hole spacetime, demonstrating the deep analogies between classical wave physics and black hole perturbation theory.

%======================================%
%<<<<<<<<<< Acknowledgement >>>>>>>>>>>%
%======================================%
\section*{Acknowledgement}
This work is supported by the National Natural Science Foundation of China. Specifically, Liang-Bi Wu is supported by Grant No. 12505067, and Zhe Yu is supported by Grant No. 12447176. This work is also supported in part by the National Key R\&D Program of China Grant No. 2022YFC2204603, by the National Natural Science Foundation of China with grants No. 12475063, No. 12075232.
\appendix
\section{The expressions of $C_{\text{in}}(\omega)$ and $C_{\text{out}}(\omega)$}\label{C_functions}
Generally, the ``in'' solution $\Psi_{\text{in}}(x)$ can be written as a linear combination of ``down'' solution $\Psi_{\text{down}}(x)$ and  ``up'' solution $\Psi_{\text{up}}(x)$, i.e., 
\begin{eqnarray}\label{Psi_in_and_Psi_down_Psi_up}
    \Psi_{\text{in}}(x)=A_{\text{in}}(\omega)\Psi_{\text{down}}(x)+A_{\text{out}}(\omega)\Psi_{\text{up}}(x)\, .
\end{eqnarray}
Letting $x\to-\infty$, one has
\begin{eqnarray}
    \mathrm{e}^{-\mathrm{i}\omega x}&=&A_{\text{in}}(\omega)\Big[C_{\text{in}}(\omega)\mathrm{e}^{-\mathrm{i}\omega x}+C_{\text{out}}(\omega)\mathrm{e}^{\mathrm{i}\omega x}\Big]+A_{\text{out}}(\omega)\Big[B_{\text{in}}(\omega)\mathrm{e}^{-\mathrm{i}\omega x}+B_{\text{out}}(\omega)\mathrm{e}^{\mathrm{i}\omega x}\Big]\nonumber\\
    &=&\Big[A_{\text{in}}(\omega)C_{\text{in}}(\omega)+A_{\text{out}}(\omega)B_{\text{in}}(\omega)\Big]\mathrm{e}^{-\mathrm{i}\omega x}+\Big[A_{\text{in}}(\omega)C_{\text{out}}(\omega)+A_{\text{out}}(\omega)B_{\text{out}}(\omega)\Big]\mathrm{e}^{\mathrm{i}\omega x}\, .
\end{eqnarray}
Comparing the coefficients of $\mathrm{e}^{-\mathrm{i}\omega x}$ and $\mathrm{e}^{\mathrm{i}\omega x}$ on both sides of the above equation, we obtain
\begin{eqnarray}
    C_{\text{in}}(\omega)=\frac{1-A_{\text{out}}(\omega)B_{\text{in}}(\omega)}{A_{\text{in}}(\omega)}\, ,\quad \text{and} \quad C_{\text{out}}(\omega)=-A_{\text{out}}(\omega)\, ,
\end{eqnarray}
where we have used the relation $A_{\text{in}}(\omega)=B_{\text{out}}(\omega)$.

\section{The numerical method of solving Eq. (\ref{master_equation_time_domain})}\label{numerical_method_TD}
In this appendix, we will give a detailed approach to solve Eq. (\ref{master_equation_time_domain}) numerically. Define $\Pi(t,x)\equiv\partial_t\Psi(t,x)$, Eq. (\ref{master_equation_time_domain}) becomes the following equations
\begin{eqnarray}\label{master_equation_time_domain_first_order_time}
    \partial_t\Psi(t,x)&=&\Pi(t,x)\, ,\nonumber\\
    \partial_t\Pi(t,x)&=&\frac{\partial^2\Psi(t,x)}{\partial x^2}-V_s(x)\Psi(t,x)\, .
\end{eqnarray}
Given the concerned interval $[x_\text{w},x_{\text{max}}]$, where $x_\text{w}\ll0$ and $x_{\text{max}}\gg0$. For $x=x_{\text{max}}$, from Eq. (\ref{boudary_condition_infinity}), we have
\begin{eqnarray}\label{boundary_condition_xmax}
    \Big(\Pi+\frac{\partial\Psi}{\partial x}\Big)\Big|_{x=x_{\text{max}}}=0\, ,\quad \text{and} \quad \Big(\frac{\partial \Pi}{\partial x}+\frac{\partial^2\Psi}{\partial x^2}\Big)\Big|_{x=x_{\text{max}}}=0\, .
\end{eqnarray}
The second equation in Eqs. (\ref{boundary_condition_xmax}) comes from the combination with the time derivative of the first equation in Eqs. (\ref{boundary_condition_xmax}) and the second equation in Eqs. (\ref{master_equation_time_domain_first_order_time}) where $V_s(x_{\text{max}})\approx0$.
For $x=x_\text{w}$, from Eq. (\ref{boundary_condition_wall}), we have
\begin{eqnarray}\label{boundary_condition_xmin}
    \Big(\mu\Pi-\frac{\partial\Psi}{\partial x}\Big)\Big|_{x=x_\text{w}}=0\, ,\quad \text{and}\quad \Big(\frac{1}{\mu}\frac{\partial \Pi}{\partial x}-\frac{\partial^2\Psi}{\partial x^2}\Big)\Big|_{x=x_\text{w}}=0\, .
\end{eqnarray}
The second equation in Eqs. (\ref{boundary_condition_xmin}) comes from the combination with the time derivative of the first equation in Eqs. (\ref{boundary_condition_xmin}) and the second equation in Eqs. (\ref{master_equation_time_domain_first_order_time}) where $V_s(x_{\text{w}})\approx0$.

One uses spectral collocation method on the interval $[x_\text{w},x_{\text{max}}]$ given the resolution $N$, where the Chebyshev-Lobatto grids
\begin{eqnarray}\label{CL_grids}
    x_{k}=\frac{x_\text{w}+x_{\text{max}}}{2}+\frac{x_{\text{max}}-x_\text{w}}{2}\cos\Big(\frac{k\pi}{N}\Big)\, ,\quad k=0,\cdots,N\, .
\end{eqnarray}
Note that $x_0=x_{\text{max}}$, $x_N=x_{\text{w}}$. Accordingly, the differential matrix $\mathbf{D}$ on the interval $[x_\text{w},x_{\text{max}}]$ is given by
\begin{eqnarray}\label{differential_matrix}
    \mathbf{D}=\frac{2}{x_{\text{max}}-x_{\text{w}}}\cdot{}^{[-1,1]}\mathbf{D}\, ,
\end{eqnarray}
where ${}^{[-1,1]}\mathbf{D}$ is standard Chebyshev-Lobatto differential matrix on the interval $[-1,1]$. For convenience, we denote that $\Psi(t,x_k)\equiv\Psi_k$ and $\Pi(t,x_k)\equiv\Pi_k$. From Eqs. (\ref{boundary_condition_xmax})-Eqs. (\ref{boundary_condition_xmin}), we obtain linear equations associated with $\Psi_0$, $\Psi_N$, $\Pi_0$ and $\Pi_N$ given by
\begin{eqnarray}\label{linear_equations}
    \mathbf{P}\cdot\begin{bmatrix}
        \Psi_0\\
        \Psi_N\\
        \Pi_0\\
        \Pi_N
    \end{bmatrix}&\equiv&
    \begin{bmatrix}
        \mathbf{D}_{00} & \mathbf{D}_{0N} & 1 & 0\\
        (\mathbf{D}^2)_{00} & (\mathbf{D}^2)_{0N} & \mathbf{D}_{00} & \mathbf{D}_{0N}\\
        -\mathbf{D}_{N0} & -\mathbf{D}_{NN} & 0 & \mu\\
        -(\mathbf{D}^2)_{N0} & -(\mathbf{D}^2)_{NN} & (1/\mu)\mathbf{D}_{N0} & (1/\mu)\mathbf{D}_{NN}
    \end{bmatrix}\cdot
    \begin{bmatrix}
        \Psi_0\\
        \Psi_N\\
        \Pi_0\\
        \Pi_N
    \end{bmatrix}\nonumber\\
    &=&
    \begin{bmatrix}
        -\sum_{i=1}^{N-1}\mathbf{D}_{0i}\Psi_i\\
        -\sum_{i=1}^{N-1}\mathbf{D}_{0i}\Pi_i-\sum_{i=1}^{N-1}(\mathbf{D}^2)_{0i}\Psi_i\\
        \sum_{i=1}^{N-1}\mathbf{D}_{Ni}\Psi_{i}\\
        -(1/\mu)\sum_{i=1}^{N-1}\mathbf{D}_{Ni}\Pi_i+\sum_{i=1}^{N-1}(\mathbf{D}^2)_{Ni}\Psi_i
    \end{bmatrix}\, .
\end{eqnarray}
Solving the above linear equations, $\Psi_0$, $\Psi_N$, $\Pi_0$ and $\Pi_N$ can be represented by the functions on internal grids namely $\Psi_1,\cdots,\Psi_{N-1}$, and $\Pi_1,\cdots,\Pi_{N-1}$. Formally, we can write them as follows 
\begin{eqnarray}\label{Psi_0_and_Psi_N}
    \Psi_0=\sum_{i=1}^{N-1}\mathbf{M}_{0i}\Psi_i+\sum_{i=1}^{N-1}\mathbf{N}_{0i}\Pi_i\, ,\quad \Psi_N=\sum_{i=1}^{N-1}\mathbf{M}_{1i}\Psi_i+\sum_{i=1}^{N-1}\mathbf{N}_{1i}\Pi_i\, ,
\end{eqnarray}
and
\begin{eqnarray}\label{Pi_0_and_Pi_N}
    \Pi_0=\sum_{i=1}^{N-1}\mathbf{M}_{2i}\Psi_i+\sum_{i=1}^{N-1}\mathbf{N}_{2i}\Pi_i\, ,\quad \Pi_N=\sum_{i=1}^{N-1}\mathbf{M}_{3i}\Psi_i+\sum_{i=1}^{N-1}\mathbf{N}_{3i}\Pi_i\, ,
\end{eqnarray}
where the dimensions of the coefficient matrices $\mathbf{M}$ and $\mathbf{N}$ are $4\times(N-1)$ and $4\times(N-1)$. The detail expressions of matrices $\mathbf{M}$ and $\mathbf{N}$ can be obtained from Eq. (\ref{linear_equations}). To be more specific, the matrix elements for the matrix $\mathbf{M}$ and $\mathbf{N}$ are written as
\begin{eqnarray}\label{matrix_M}
    \mathbf{M}_{ki}=-(\mathbf{P}^{-1})_{k0}\mathbf{D}_{0i}-(\mathbf{P}^{-1})_{k1}(\mathbf{D}^2)_{0i}+(\mathbf{P}^{-1})_{k2}\mathbf{D}_{Ni}+(\mathbf{P}^{-1})_{k3}(\mathbf{D}^2)_{Ni}\, ,
\end{eqnarray}
and
\begin{eqnarray}\label{matrix_N}
    \mathbf{N}_{ki}=-(\mathbf{P}^{-1})_{k1}\mathbf{D}_{0i}-(1/\mu)(\mathbf{P}^{-1})_{k3}\mathbf{D}_{Ni}\, ,
\end{eqnarray}
for $k=0,1,2,3$ and $i=1,2,\cdots, N-1$. $\mathbf{P}^{-1}$ denotes the inverse of the matrix $\mathbf{P}$. It is worth mentioning that the elements of matrix $\mathbf{M}$ and matrix $\mathbf{N}$ are labeled with rows $0$ to $3$ and columns $1$ to $N-1$.

Now, from the Chebyshev-Lobatto grid (\ref{CL_grids}), Eqs. (\ref{master_equation_time_domain_first_order_time}) can be discretized into the following ODEs
\begin{eqnarray}\label{master_equation_time_domain_discretize}
    \frac{\mathrm{d}\Psi_i}{\mathrm{d} t}&=&\Pi_i\, ,\nonumber\\
    \frac{\mathrm{d}\Pi_i}{\mathrm{d} t}&=&\sum_{j=1}^{N-1}(\mathbf{D}^2)_{ij}\Psi_j+(\mathbf{D}^2)_{i0}\Psi_0+(\mathbf{D}^2)_{iN}\Psi_N-V_s(x_i)\Psi_i\nonumber\\
    &=&\sum_{j=1}^{N-1}(\mathbf{D}^2)_{ij}\Psi_j+(\mathbf{D}^2)_{i0}\Big(\sum_{j=1}^{N-1}\mathbf{M}_{0j}\Psi_j+\sum_{j=1}^{N-1}\mathbf{N}_{0j}\Pi_j\Big)+(\mathbf{D}^2)_{iN}\Big(\sum_{j=1}^{N-1}\mathbf{M}_{1j}\Psi_j+\sum_{j=1}^{N-1}\mathbf{N}_{1j}\Pi_j\Big)\nonumber\\
    &&-V_s(x_i)\Psi_i\, ,
\end{eqnarray}
for $i=1,\cdots,N-1$.  Finally, we have the following ODEs:
\begin{eqnarray}\label{odes_metrix_version}
    \frac{\mathrm{d}}{\mathrm{d}t}
    \begin{bmatrix}
       \Psi_1\\
    \vdots\\
    \Psi_{N-1}\\
    \Pi_1\\
    \vdots\\
    \Pi_{N-1} 
    \end{bmatrix}
    =\begin{bmatrix}
        \mathbf{O} & \mathbf{I}\\
        \mathbf{L}_1 & \mathbf{L}_2
    \end{bmatrix}\cdot
    \begin{bmatrix}
       \Psi_1\\
    \vdots\\
    \Psi_{N-1}\\
    \Pi_1\\
    \vdots\\
    \Pi_{N-1} 
    \end{bmatrix}
    \equiv\mathbf{L}\cdot
    \begin{bmatrix}
       \Psi_1\\
    \vdots\\
    \Psi_{N-1}\\
    \Pi_1\\
    \vdots\\
    \Pi_{N-1} 
    \end{bmatrix}\, ,
\end{eqnarray}
where the matrix $\mathbf{L}$ is obtained from Eq. (\ref{master_equation_time_domain_discretize}), and it is independent of the time. The dimensions of $\mathbf{L}_1$ and $\mathbf{L}_2$ are $(N-1)\times(N-1)$ and $(N-1)\times(N-1)$. Furthermore, the matrix elements for $\mathbf{L}_1$ and $\mathbf{L}_2$ read
\begin{eqnarray}
    (\mathbf{L}_{1})_{ij}=(\mathbf{D}^2)_{ij}+(\mathbf{D}^2)_{i0}\mathbf{M}_{0j}+(\mathbf{D}^2)_{iN}\mathbf{M}_{1j}-\mathbf{V}_{ij}\, ,\quad i=1,\cdots,N-1\, ,\quad j=1,\cdots,N-1\, ,
\end{eqnarray}
where the potential matrix $\mathbf{V}_{ij}$ is
\begin{eqnarray}
    \mathbf{V}=\text{diag}\Big\{V_s(x_0),\cdots,V_s(x_{N})\Big\}\, ,
\end{eqnarray}
and
\begin{eqnarray}
    (\mathbf{L}_2)_{ij}=(\mathbf{D}^2)_{i0}\mathbf{N}_{0j}+(\mathbf{D}^2)_{iN}\mathbf{N}_{1j}\, ,\quad i=1,\cdots,N-1\, ,\quad j=1,\cdots,N-1\, .
\end{eqnarray}

To solve such ODEs (\ref{odes_metrix_version}) numerically, we adopt a discrete time evolution scheme via a $6$th-order Hermite integration method~\cite{markakis2019timesymmetry,OBoyle:2022yhp,Markakis:2023pfh,DaSilva:2024yea,Cao:2024sot,Guo:2026zjq,He:2025ydh,Cao:2025qws}, which is actually an implicit scheme. Selecting a constant time step $\Delta t$, we arrive at
\begin{eqnarray}\label{Solution_u}
    \mathbf{u}\Big((i+1)\Delta t\Big) = \mathbf{U}\cdot\mathbf{u}(i\Delta t),\quad i=0,1,\cdots\, ,
\end{eqnarray}
where $\mathbf{U}$ is called evolution matrix whose explicit form can be expressed as
\begin{eqnarray}\label{evolution_matrix}
    \mathbf{U}=\mathbf{I}+(\Delta t \mathbf{L})\cdot\Big[\mathbf{I}+\frac{1}{60}(\Delta t \mathbf{L})\cdot(\Delta t \mathbf{L})\Big]\cdot\Bigg\{\mathbf{I}-\frac{\Delta t}2\mathbf{L}\cdot\Big[\mathbf{I}-\frac{\Delta t}5\mathbf{L}\cdot\Big(\mathbf{I}-\frac{\Delta t}{12}\mathbf{L}\Big)\Big]\Bigg\}^{-1},
\end{eqnarray}
and $\mathbf{I}$ is the identity matrix whose dimension is consistent with that of $\mathbf{L}$. Because such scheme is unconditionally stable, there is no Courant limit on the time step $\Delta t$. Meanwhile, there is no requirement for constant time steps, i.e., $\Delta t$ can depend on $i$. For convenience, we simply take even time steps in our practice. In addition, because Eq. (\ref{Solution_u}) involves the numerical operation of constantly multiplying a vector by the same matrix, we use the built-in function \textit{NestList} in \textit{Mathematica} to accelerate this process.

\section{Integral on subinterval for the Chebyshev-Lobatto grids}\label{Integral_method}
Similar to the Ref. \cite{Tuncer:2025dnp}, the energy integral on the subinterval is necessary. Hence, for the convenience of numerical calculations, we provide the integral expression on the subinterval within the framework of Chebyshev-Lobatto grids. Our goal is to get the expression of the following integral
\begin{eqnarray}\label{E_ab}
    E_{ab}(t)=\int_{x_a}^{x_b}\rho(t,x)\mathrm{d}x\, ,
\end{eqnarray}
given some time $t$, where $x_a$ and $x_b$ are satisfied with $x_{\text{w}}\leq x_{a}< x_b\leq x_{\text{max}}$. Define a new variable $y$ via
\begin{eqnarray}
    x\equiv\frac{x_\text{w}+x_{\text{max}}}{2}+\frac{x_{\text{max}}-x_\text{w}}{2}y\, ,\quad y\in[-1,1]\, ,
\end{eqnarray}
the above integral (\ref{E_ab}) becomes
\begin{eqnarray}
    E_{ab}(t)=\frac{x_{\text{max}}-x_\text{w}}{2}\int_{y_a}^{y_b}\rho\Big(t,\frac{x_\text{w}+x_{\text{max}}}{2}+\frac{x_{\text{max}}-x_\text{w}}{2}y\Big)\mathrm{d}y\equiv\frac{x_{\text{max}}-x_\text{w}}{2}\int_{y_a}^{y_b}g(y)\mathrm{d}y\, ,
\end{eqnarray}
where $y_a$ and $y_b$ are satisfied with $-1\leq y_{a}< y_b\leq 1$. Then, $E_{ab}(t)$ can be approximated by using the Chebyshev polynomials, namely
\begin{eqnarray}
    E_{ab}(t)\approx\frac{x_{\text{max}}-x_\text{w}}{2}\int_{y_a}^{y_b}\sum_{k=0}^{N}a_kT_k(y)\mathrm{d}y=\frac{x_{\text{max}}-x_\text{w}}{2}\sum_{k=0}^{N}a_kJ_k\, ,
\end{eqnarray}
where the Chebyshev polynomial expansion coefficients $a_k$'s are given by
\begin{eqnarray}
    a_k=\frac{2}{Nc_k}\sum_{j=0}^N\frac{g_j}{c_j}\cos\Big(\frac{jk\pi}{N}\Big)\, ,\quad k=0,\cdots,N\, ,\quad \text{with}\quad c_0=c_N=2\, ,\quad c_k=1 (k=1,\cdots,N-1)\, ,
\end{eqnarray}
and $J_0=y_b-y_a$, $J_1=(y_b^2-y_a^2)/2$,
\begin{eqnarray}
    J_k=\int_{y_a}^{y_b}T_k(y)\mathrm{d}y=\frac{1}{2}\Big[\frac{T_{k+1}(y_b)-T_{k+1}(y_a)}{k+1}-\frac{T_{k-1}(y_b)-T_{k-1}(y_a)}{k-1}\Big]\, ,\quad k\geq2\, .
\end{eqnarray}
Note that $g_j=\rho_j=\rho(t,x_j)$, $j=0,\cdots,N$. Finally, we have
\begin{eqnarray}\label{Eab_t}
    E_{ab}(t)\approx\sum_{k=0}^Nw_k\rho_k\, ,\quad w_k=\frac{x_{\text{max}}-x_\text{w}}{2}\cdot \frac{2}{Nc_k}\sum_{j=0}^N\frac{J_j}{c_j}\cos\Big(\frac{jk\pi}{N}\Big)\, .
\end{eqnarray}
Because $E_{ab}(t)$ depends on the values of $\rho$ at the grid points, including the values at the endpoints of the interval, we need to use Eqs. (\ref{Psi_0_and_Psi_N}) and Eqs. (\ref{Pi_0_and_Pi_N}) to obtain the values of $\Psi$ and $\Pi$ at the endpoints of the interval, thereby obtaining the values of $\rho$ at the endpoints.

\section{Numerical resolution test}\label{Numerical_convergence_test}
In this appendix, we perform numerical resolution tests to examine the dependence of the results on the spatial and temporal resolutions. For the parameters with $r_\text{c}=1.1$, $x_{\text{w}}=-100$, $\mathcal{K}=0.03162$, $\mathcal{A}=1$, $x_{\text{t}}=800$ and $\sigma=30$ (Fig. \ref{Energy_Plot_rc_1p1}), numerical results of $\eta$ are shown in Tab. \ref{numerical_test_rc_1p1}. For the parameters with $r_\text{c}=2$, $x_{\text{w}}=-20$, $\mathcal{K}=-0.09261$, $\mathcal{A}=1$, $x_{\text{t}}=600$ and $\sigma=18$ (Fig. \ref{Energy_Plot_rc_2}), numerical results of $\eta$ are shown in Tab. \ref{numerical_test_rc_2}. From Tables \ref{numerical_test_rc_1p1} and \ref{numerical_test_rc_2}, the difference in $\eta$ values between simulations of different resolutions is relatively small. The numerical results are stable against changes in the spatial and temporal resolutions used in our simulations.
    
     \begin{table}[ht]
\setlength{\tabcolsep}{12pt} 
 \centering
\begin{tabular}{c c c c c}
 \hline\hline
Numerical settings~~~~~~& $\eta(\omega=\omega_{\text{VAM}})$ ~~~~~~& $\eta(\omega=0.9\omega_{\text{VAM}})$ ~~~~~~&$\eta(\omega=0.6\omega_{\text{VAM}})$ ~~~~~~&  \\
 \hline\noalign{\vskip 5pt}
$N=330$, $\mathrm{d}t=0.1$~~~~~~& $1.368\times10^{-4}$ ~~~~~~& $4.937\times10^{-4}$ ~~~~~~&$5.160\times10^{-2}$ ~~~~~~&  \\\noalign{\vskip 5pt}
 \hline\noalign{\vskip 5pt}
$N=340$, $\mathrm{d}t=0.1$~~~~~~& $1.465\times10^{-4}$ ~~~~~~& $5.097\times10^{-4}$ ~~~~~~&$5.159\times10^{-2}$ ~~~~~~&  \\\noalign{\vskip 5pt}
\hline\noalign{\vskip 5pt}
$N=350$, $\mathrm{d}t=0.2$~~~~~~& $1.564\times10^{-4}$ ~~~~~~& $5.254\times10^{-4}$ ~~~~~~&$5.158\times10^{-2}$ ~~~~~~&  \\\noalign{\vskip 5pt}
\hline\noalign{\vskip 5pt}
$N=350$, $\mathrm{d}t=0.1$~~~~~~& $1.564\times10^{-4}$ ~~~~~~& $5.254\times10^{-4}$ ~~~~~~&$5.158\times10^{-2}$ ~~~~~~&\\\noalign{\vskip 5pt}
\hline\noalign{\vskip 5pt}
$N=350$, $\mathrm{d}t=0.05$~~~~~~& $1.564\times10^{-4}$ ~~~~~~& $5.254\times10^{-4}$ ~~~~~~&$5.158\times10^{-2}$ ~~~~~~&  \\\noalign{\vskip 5pt}
\hline\hline
\end{tabular}
\caption{The numerical resolution tests for the parameters with $r_\text{c}=1.1$, $x_{\text{w}}=-100$, $\mathcal{K}=0.03162$, $\mathcal{A}=1$, $x_{\text{t}}=800$ and $\sigma=30$.}
\label{numerical_test_rc_1p1}
\end{table}

\begin{table}[ht]
\setlength{\tabcolsep}{12pt} 
 \centering
\begin{tabular}{c c c c c}
 \hline\hline
Numerical settings~~~~~~& $\eta(\omega=\omega_{\text{VAM}})$ ~~~~~~& $\eta(\omega=0.9\omega_{\text{VAM}})$ ~~~~~~&$\eta(\omega=0.6\omega_{\text{VAM}})$ ~~~~~~&  \\
 \hline\noalign{\vskip 5pt}
$N=370$, $\mathrm{d}t=0.1$~~~~~~& $3.182\times10^{-3}$ ~~~~~~& $1.552\times10^{-2}$ ~~~~~~&$0.5888$ ~~~~~~&    \\\noalign{\vskip 5pt}
 \hline\noalign{\vskip 5pt}
$N=380$, $\mathrm{d}t=0.1$~~~~~~& $3.262\times10^{-3}$ ~~~~~~& $1.546\times10^{-2}$ ~~~~~~&$0.5888$ ~~~~~~&  \\\noalign{\vskip 5pt}
\hline\noalign{\vskip 5pt}
$N=390$, $\mathrm{d}t=0.2$~~~~~~& $3.290\times10^{-3}$ ~~~~~~& $1.551\times10^{-2}$ ~~~~~~&$0.5888$ ~~~~~~&   \\\noalign{\vskip 5pt}
\hline\noalign{\vskip 5pt}
$N=390$, $\mathrm{d}t=0.1$~~~~~~& $3.290\times10^{-3}$ ~~~~~~& $1.551\times10^{-2}$ ~~~~~~&$0.5888$ ~~~~~~&  \\\noalign{\vskip 5pt}
\hline\noalign{\vskip 5pt}
$N=390$, $\mathrm{d}t=0.05$~~~~~~& $3.290\times10^{-3}$ ~~~~~~& $1.551\times10^{-2}$ ~~~~~~&$0.5888$ ~~~~~~&  \\\noalign{\vskip 5pt}
\hline\hline
\end{tabular}
\caption{The numerical resolution tests for the parameters with $r_\text{c}=2$, $x_{\text{w}}=-20$, $\mathcal{K}=-0.09261$, $\mathcal{A}=1$, $x_{\text{t}}=600$ and $\sigma=18$.}
\label{numerical_test_rc_2}
\end{table}

\bibliography{reference}
\bibliographystyle{apsrev4-1}

\end{document}